\documentclass[draftcls, onecolumn, 12pt]{IEEEtran}
\usepackage{fancyhdr}
\usepackage{flushend}
\usepackage{threeparttable}
\usepackage{epsf,epsfig}
\usepackage{amsthm}
\usepackage{amsmath}
\usepackage{amssymb}
\usepackage{amsfonts}
\usepackage[noadjust]{cite}
\usepackage{mcite}
\usepackage{dsfont}
\usepackage{subfigure}
\usepackage{color}
\usepackage{algorithm}
\usepackage{algpseudocode}
\usepackage{enumerate}
\usepackage{graphicx}

\begin{document}

\title{Generalized Distributed \\Compressive Sensing}

\author{Jeonghun~Park~\IEEEmembership{Student Member,~IEEE,}
        Seunggye~Hwang~\IEEEmembership{Student Member,~IEEE,}
        Janghoon~Yang,~\IEEEmembership{Member,~IEEE,}
        and~Dongku~Kim*,~\IEEEmembership{Member,~IEEE}
\thanks{J.-H Park, S.-K Hwang and D.-K Kim (*corresponding author) are with the School of Electrical and Electronic Engineering, Yonsei University, Seoul, Korea. Emails:${\rm{\{ the20thboys, pisces\_sg, dkkim\} }}$@yonsei.ac.kr, Tel: 82-2-2123-2877, Fax: 82-2-365-4504.
J.-H Yang is with Department of Newmedia, Korean German Institute of Technology, Seoul, Korea. Email:jhyang@kgit.ac.kr, Tel: 82-2-6393-3237.}}

\maketitle

\begin{abstract}

Distributed Compressive Sensing (DCS) \cite{Baron:2009vd} 
improves the signal recovery performance of multi signal
ensembles by exploiting both intra- and inter-signal correlation and sparsity structure. 
However, the existing DCS was proposed for a very 
limited ensemble of signals that has single common information \cite{Baron:2009vd}.
In this paper, we propose a generalized DCS (GDCS) which can improve sparse signal detection performance
given arbitrary types of common information
which are classified into not just full common information but also a variety of partial common information.
The theoretical bound on the required number of measurements using the GDCS
is obtained. Unfortunately, the GDCS may require much a priori-knowledge on 
various inter common information of ensemble of signals to enhance the performance over the 
existing DCS.
To deal with this problem, we propose a novel algorithm 
that can search for the correlation structure among the signals,
with which the proposed GDCS improves 
detection performance even
without a priori-knowledge on correlation structure
for the case of arbitrarily correlated multi signal ensembles.

\end{abstract}


\IEEEpeerreviewmaketitle

\section{Introduction}

Generally, signals in various applications can be represented as sparse coefficients with a particular basis, meaning 
a signal vector ${\bf{x}} \in {\mathbb{R}^N}$
has only $K \ll N$ nonzero coefficients. Many compression algorithms exploit 
this sparse structure, including MP3 \cite{brandenburg1999mp3}, JPEG \cite{Pennebaker:1992:JSI:573326} and JPEG2000 
\cite{Taubman:2001:JIC:559856}.
Compressive sensing (CS) is an emerging signal acquisition technique that has an
advantage of reducing the required number of measurements for recovery of sparse signal. If a target signal ${\bf{x}} \in {\mathbb{R}^N}$ is represented as a sparse signal with a
particular sparse basis, one can recover it with only $M < N$ measurements. 
It is known that 
the 
signal can be recovered with overwhelming probability if
the sparsity $K$ (simply, the number of nonzero elements) of the signal satisfies
$K \le C{\left( {\log N} \right)^{ - 1}}M$ \cite{Candes:2006eq},
where $C$ is a constant.

Baron \emph{et al.} \cite{Baron:2009vd} introduced Distributed Compressive Sensing (DCS),
which exploits not just intra-, but also inter- joint sparsity to improve the detection performance.
They assume the scenario of
a Wireless Sensor Network (WSN) 
consisting of an arbitrary number of sensors and one sink node.
In this scenario, each sensor should normally carry out the compression in a distributed way without
cooperation of the other sensors and transmit the compressed signal to the sink node.
At the sink node
the received signals from all the sensors are reconstructed jointly. 
Here, a key of the DCS is the concept of joint sparsity, defined as the sparsity of the entire signal 
ensemble. 
Three models have been considered
as joint sparse signal models in \cite{Baron:2009vd}.
In the first model, 
not only each signal is individually sparse, but there are also 
common components shared by every signal, called common information,
which allow reduction of required measurements by joint recovery.
In the second model, all signals share the supports, the locations of the nonzero coefficients.
In the third model,
no signal is sparse itself, nevertheless, 
they share the large amount of common information, which makes it possible to compress and recover the signals.
While the second model, called
the Multiple Measurement Vector (MMV) setting, has been actively explored in \cite{{Tropp:2005du},{Davies:2012jv},{Kim:2012kw}},
to the best of authors' knowledge,
the first model has been studied for only a limited ensemble of signals that has single common information.

Despite its limitation, 
the first model is applied in many other applications such as \cite{{Zhang:2010bv},{Yin:2011em}} 
as well as in the WSN \cite{Baron:2009vd}.
In \cite{Zhang:2010bv}, they extract a common component and an innovation component from various face images 
for facilitating an analysis task such as face recognition.
In \cite{Yin:2011em}, when implementing image fusion that combines multiple images of the same scene
into a single image which is suitable for human perception and practical applications, 
they model the constant background image as common information 
and the variable foreground image as innovation information for efficiency of the process.

However,
it is unrealistic to assume that there exists
only common information.
Practically, in most of situations, partial common information, which is firstly 
proposed in our conference version paper \cite{Park:vn}, as well as full common information are 
measured by arbitrary number of multiple sensors.
Using this notion, we 
introduce partial common information 
leading to a generalized DCS (GDCS) model in this paper
and obtain the theoretical bound of the number of measurements
for exact reconstruction.
However, 
to take advantage of partial common information,
the decoder should know partial common structures of signals, 
which is not typically known to the decoder.
To deal with this problem, we also propose a novel algorithm that can find
the correlation structure among sensors to help
decoder exploit partial common information.
This algorithm can provide significant performance improvement.
In summary, the main contributions of this paper are as follows.

\begin{enumerate} [$\circ $]

\item We propose a GDCS model where \cite{Baron:2009vd} is a special case.

\item The theoretical bound on the required number of measurements of the GDCS model is obtained.

\item To solve the necessity of a priori-knowledge, which is a burden of the decoder, we propose a novel algorithm that iteratively detects the signals with the proposed algorithm.

\end{enumerate}


The remainder of this paper is organized as follows.
We summarize the background of CS briefly in Section II. In Section III, 
we explain the concept
of the existing joint sparse signal model and define its general version extension. 
Based on this model, we obtain the theoretical 
bound on the required number of measurements and propose a novel algorithm to 
capitalize on the GDCS in a practical environment in Section VI. 
In Section VII, numerical simulations are provided, followed
by conclusions in Section VIII.

\section{Compressive sensing background}
When we deal with the signals sensed in the real world, in many cases
we can represent a real value signal ${\bf{x}} \in {\mathbb{R}^N}$ as sparse coefficients with a particular basis 
${\bf{\Psi }} = \left[ {{\psi _1},...,{\psi _N}} \right]$. We can write
\begin{equation} \label{eq1}
{\bf{x}} = \sum\limits_{n = 1}^N {{\psi _n}\omega \left( n \right)} 
\end{equation}
where $\omega \left( n \right)$ is the $n$th component of sparse coefficients $\omega $ and
${\psi _n}$ is the $n$th column of the sparse basis.
Without loss of generality, let assume that ${\left\| \omega  \right\|_0} = K$.
Here, ${\left\| \omega  \right\|_0}$ is the number of nonzero elements in vector $\omega $.
In matrix multiplication form, this is represented as
\begin{equation} \label{eq2}
{\bf{x}} = {\bf{\Psi }}\omega 
\end{equation}
Including the widely used Fourier and wavelet basis, various expansions, e.g., Gabor bases \cite{Mallat:2008:WTS:1525499}
and bases obtained by Principal Component Analysis (PCA) \cite{Masiero:2009ft}, can be used  
as a sparse basis. For convenience, we use the identity matrix ${\bf{I}}$ for a sparse basis $\bf{\Psi} $. Without loss of 
generality, an arbitrary sparse basis is easily incorporated into the developed structure.

Candes, Romberg and Tao \cite{Candes:2006eq} and Donoho \cite{Donoho:2006ci} showed that a reduced set of linear projections can contain enough information
to recover the sparse signal. This technique introduced in \cite{{Candes:2006eq},{Donoho:2006ci}} has been named CS. 
In the CS, a compression is processed by simply projecting a signal onto measurement matrix ${\bf{\Phi }} \in {\mathbb{R}^{M \times N}}$ where $M \ll N$.
We can describe the compression procedure as follows. 
\begin{equation} \label{eq3}
\begin{array}{l}
{\bf{y}} = {\bf{\Phi x}}\\
{\rm{where}}\;\;{\bf{y}} \in {\mathbb{R}^M}
\end{array}
\end{equation}
Since the number of equations $\left( M \right)$ is smaller than the number of values $\left( N \right)$, this system is ill-posed.
However, the sparsity of the signal allows perfect recovery
if 
the restricted isometry property (RIP) of $\Phi $ \cite{Candes:2006eq}, \cite{Candes:2005cs} is satisfied with an appropriate constant. 
Assuming that the signal ${\bf{x}}$ can be represented as in (\ref{eq2}), the sparsest coefficient vector $\hat \omega$ 
can be found by solving the following $l_0$ minimization.
\begin{equation} \label{eq4}
\hat \omega  = \arg \min {\left\| \omega  \right\|_0}\;\;\;\;s.t.\;{\bf{y}} = {\bf{\Phi \Psi }}\omega
\end{equation}
If the original coefficient vector $\omega$ 
is sparse enough, there is no other
sparse solution that satisfies ${\bf{y}} = {\bf{\Phi \Psi }}\hat \omega $ except for $\hat \omega = \omega$,
which implies we can recover the original signal in spite of the ill-posedness of the system.

However, although ${l_0}$ minimization problem guarantees significant 
reduction in the required number of measurements for recovery,
we cannot use ${l_0}$ minimization practically because of its huge complexity. To solve the ${l_0}$ minimization, we must search
${\left( {\begin{array}{*{20}{c}}
N\\
K
\end{array}} \right)} $ possible sparse subspaces, which makes ${l_0}$ minimization NP-hard \cite{DBLP:conf/focs/CandesRTV05}. 

Instead of solving ${l_0}$ minimization, 
we can use the solution of ${l_1}$ minimization as the coefficient vector of the original signal, paying more measurements \cite{Candes:2006eq} as a cost of a tractable algorithm.
\begin{equation} \label{eq5}
\hat \omega  = \arg \min {\left\| \omega  \right\|_1}\;\;\;\;s.t.\;{\bf{y}} = {\bf{\Phi \Psi }}\omega 
\end{equation}
This approach is called Basis Pursuit. Contrary to ${l_0}$ minimization, we can solve ${l_1}$ minimization with 
bearable complexities, which is polynomial in $N$.

Not only norm minimization, but also an iterative greedy algorithm can be used for finding the original signal from
the observed signal. The Orthogonal Matching Pursuit (OMP) \cite{citeulike:4109428} is the most typical algorithm among iterative greedy algorithms. 
It iteratively chooses the vector from the measurement matrix ${\bf{\Phi \Psi }}$ that occupies the largest portion in the observed signal ${\bf{y}}$.
It is proven in \cite{citeulike:4109428} that
the original signal can be recovered with appropriately high probability by OMP. A greedy algorithm
has been developed to more sophisticated algorithm, 
e.g., CoSaMP \cite{Needell:2009gp} and Subspace Pursuit \cite{Dai:2009hg}.

\section{Joint Sparse Signal Model}

In \cite{Baron:2009vd}, the joint sparse signal model is defined. Using the same notations with \cite{Baron:2009vd}, let $\Lambda : = \left\{ {1,2,...,J} \right\}$ 
denote the set of indices of signal ensembles.
The ensembles consist of the signal ${{\bf{x}}_j} \in {\mathbb{R}^N},\;\;j \in \Lambda $. 
We use ${{\bf{x}}_j}\left( n \right)$ as the $n$th sample in the signal $j$.
Each sensor $j$ is given a distinct measurement matrix ${{\bf{\Phi }}_j} \in {{\mathbb{R}^{{M_j} \times N}}}$,
which is i.i.d. Gaussian matrix.
The compressed signal ${{\bf{y}}_j} \in {{\mathbb{R}^{{M_j}}}}$  can be written as ${{\bf{y}}_j} = {{\bf{\Phi }}_j}{{\bf{x}}_j}$. Concatenating all the signals from 1 to $J$, we can write it in the following form.
\begin{equation} \label{eq6}
{\bf{X}} = \left[ {\begin{array}{*{20}{c}}
{{{\bf{x}}_1}}\\
{{{\bf{x}}_2}}\\
 \vdots \\
{{{\bf{x}}_J}}
\end{array}} \right],\;\;\;{\bf{Y}} = \left[ {\begin{array}{*{20}{c}}
{{{\bf{y}}_1}}\\
{{{\bf{y}}_2}}\\
 \vdots \\
{{{\bf{y}}_J}}
\end{array}} \right],\;\;\;{\rm{and}}\;\;{\bf{\Phi }} = \left[ {\begin{array}{*{20}{c}}
{{{\bf{\Phi }}_1}}&0& \cdots &0\\
0&{{{\bf{\Phi }}_2}}& \cdots &0\\
 \vdots & \vdots & \ddots & \vdots \\
0&0& \cdots &{{{\bf{\Phi }}_J}}
\end{array}} \right]
\end{equation}
where ${\bf{X}} \in {\mathbb{R}^{JN}}$, ${\bf{Y}} \in {\mathbb{R}^{\sum\limits_{j \in \Lambda } {{M_j}} }}$ and ${\bf{\Phi }} \in {\mathbb{R}^{\sum\limits_{j \in \Lambda } {{M_j}}  \times JN}}$
. Finally we can write
\begin{equation} \label{eq7}
{\bf{Y}} = {\bf{\Phi X}}
\end{equation}

In \cite{Baron:2009vd}, ${{\bf{x}}_j}$ can be decomposed into two parts. 
The first part is common information ${{\bf{z}}_C}$, which is measured by every sensor, and
the second part is innovation information ${{\bf{z}}_j},\;j \in \Lambda $, which is uniquely measurable by the sensor $j$.
The signal ${{\bf{x}}_j}$ can be written accordingly as 
\begin{equation} \label{eq8}
{{\bf{x}}_j} = {{\bf{z}}_C} + {{\bf{z}}_j},\;\;j \in \Lambda 
\end{equation}

While (\ref{eq8}) is composed of two kinds of components,
we can refine the model by defining 
the partial common information as follows : 
It is the information measured by $\rho $ multiple sensors where 
$\rho $ is an arbitrary number that satisfies $1 < \rho  < J$.
Then, the innovation information in (\ref{eq8})
can be decomposed 
into partial common/innovation information.
\begin{figure} [t]
\centering
\includegraphics [width=3.5in,height=2.4in]  {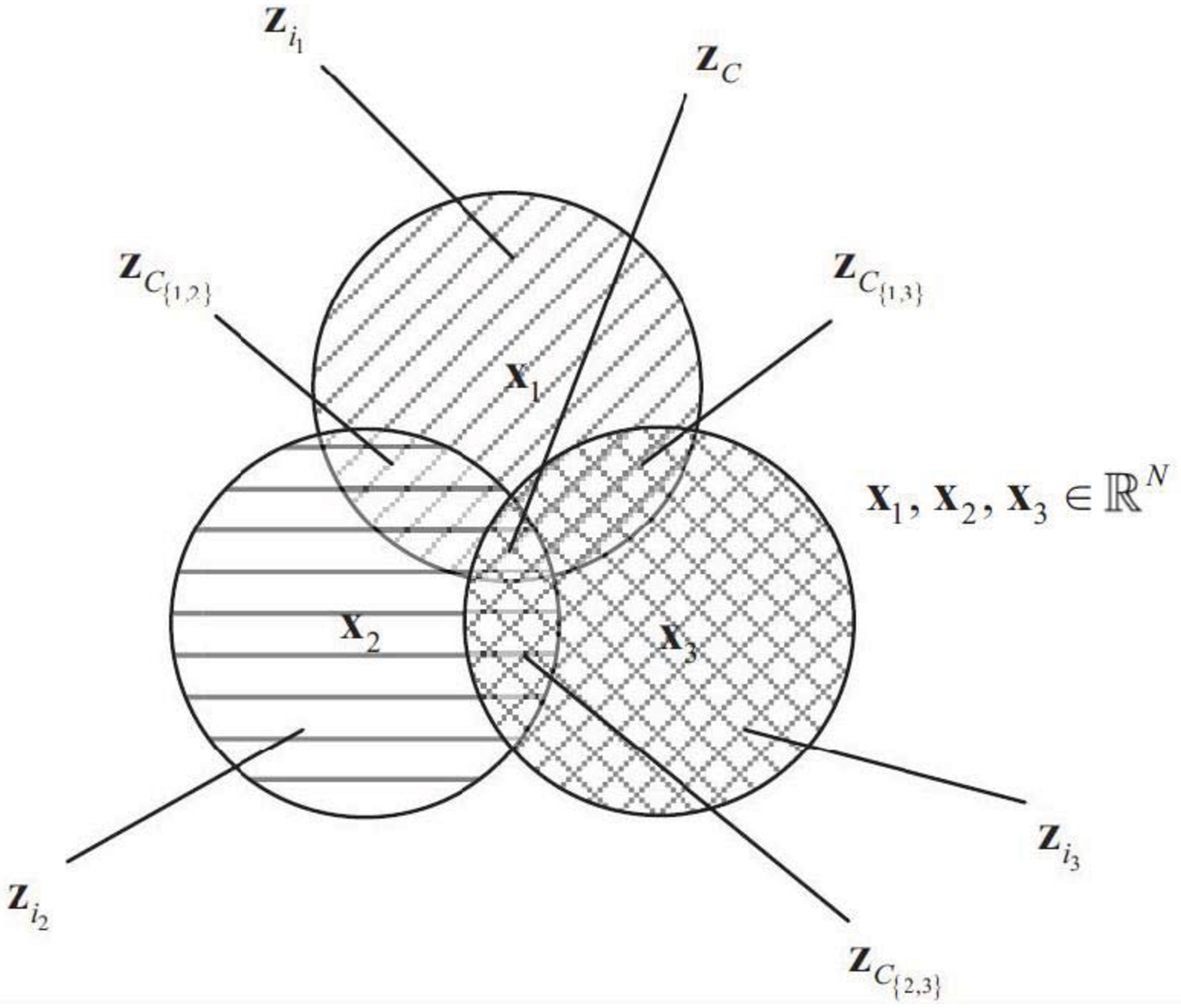}
\caption{Venn diagram description of the characterizing signal model for the DCS} \label{fig1}
\end{figure}
For ease of explanation, we consider a simple sensor network in Fig. \ref{fig1}
where each sensor measures its own innovation information and the full common information of all three sensors.
In addition to those, three different partial common information 
can be measured in pairs as \{sensor 1, sensor 2\}, \{sensor 2, sensor 3\}, and \{sensor 3 and sensor 1\}. 
For ${j_1},{j_2} \in \Lambda $ and ${j_1} \ne {j_2}$, let ${{\bf{z}}_{{C_{\left\{ {{j_1},{j_2}} \right\}}}}}$ denote
partial common information measured by 
the sensors $\left\{ {{j_1},\;{j_2}} \right\}$.
To avoid confusion with the notation of the existing signal model, we change the notation of innovation information to ${{\bf{z}}_{{i_j}}},\;j \in \Lambda $
in our model. 
With the defined notion, we can write the signal ${{\bf{x}}_j}$ of Fig. \ref{fig1} in the following form.
\begin{equation} \label{eq9}
\begin{array}{l}
{{\bf{x}}_1} = {{\bf{z}}_C} + {{\bf{z}}_{{C_{\left\{ {1,2} \right\}}}}} + {{\bf{z}}_{{C_{\left\{ {1,3} \right\}}}}} + {{\bf{z}}_{{i_1}}}\\
{{\bf{x}}_2} = {{\bf{z}}_C} + {{\bf{z}}_{{C_{\left\{ {2,1} \right\}}}}} + {{\bf{z}}_{{C_{\left\{ {2,3} \right\}}}}} + {{\bf{z}}_{{i_2}}}\\
{{\bf{x}}_3} = {{\bf{z}}_C} + {{\bf{z}}_{{C_{\left\{ {3,2} \right\}}}}} + {{\bf{z}}_{{C_{\left\{ {3,1} \right\}}}}} + {{\bf{z}}_{{i_3}}}
\end{array}
\end{equation}
We can readily extend (\ref{eq9}) to the case of an arbitrary large number of sensors.

%

Adopting the same notation in DCS \cite{Baron:2009vd} which decouples the location and value of a signal,
we also write an arbitrary sparse signal ${\bf{x}}$ as
\begin{equation} \label{form2}
{\bf{x}} = P\theta 
\end{equation}
for ${\bf{x}} \in {\mathbb{R}^N}$ satisfying ${\left\| {\bf{x}} \right\|_0} = K$, where
$\theta  \in {\mathbb{R}^K}$, called a value vector, contains only nonzero elements in ${\bf{x}}$, and 
$P \in {\mathbb{R}^{N \times K}}$, called a location matrix, is an identity submatrix, 
which consists of $K$ column vectors chosen from an $N \times N$ identity matrix. 
With these, we can describe the signal model (\ref{eq9}) as follows.


\begin{equation} \label{eq11}
\begin{array}{l}
{\bf{X}} = {\left[ {\begin{array}{*{20}{c}}
{{\bf{x}}_1^T}& \cdots &{{\bf{x}}_3^T}
\end{array}} \right]^T} \in {\mathbb{R}^{3N}}\\
{\bf{X}} = {\bf{P}}\Theta ,\;\;{\rm{where}}\\
{\bf{P}} \in {\mathbb{R}^{3N \times \left( {{K_C}\left( {\bf{P}} \right) + \sum\limits_{{j_1},\;{j_2},\;{j_1} \ne {j_2}} {{K_{{C_{\left\{ {{j_1},{j_2}} \right\}}}}}\left( {\bf{P}} \right)}  + \sum\limits_{j = 1}^3 {{K_{{i_j}}}\left( {\bf{P}} \right)} } \right)}}\;\;\\
\Theta  \in {\mathbb{R}^{\left( {{K_C}\left( {\bf{P}} \right) + \sum\limits_{{j_1},\;{j_2},\;{j_1} \ne {j_2}} {{K_{{C_{\left\{ {{j_1},{j_2}} \right\}}}}}\left( {\bf{P}} \right)}  + \sum\limits_{j = 1}^3 {{K_{{i_j}}}\left( {\bf{P}} \right)} } \right)}}\\
{\bf{P}} = \left[ {\begin{array}{*{20}{c}}
{{P_C}}&{{P_{{C_{\left\{ {1,2} \right\}}}}}}&{{P_{{C_{\left\{ {1,3} \right\}}}}}}&0&{{P_{{i_1}}}}&0&0\\
{{P_C}}&{{P_{{C_{\left\{ {1,2} \right\}}}}}}&0&{{P_{{C_{\left\{ {2,3} \right\}}}}}}&0&{{P_{{i_2}}}}&0\\
{{P_C}}&0&{{P_{{C_{\left\{ {1,3} \right\}}}}}}&{{P_{{C_{\left\{ {2,3} \right\}}}}}}&0&0&{{P_{{i_3}}}}
\end{array}} \right]\;\;\\
\Theta  = {\left[ {\begin{array}{*{20}{c}}
{\theta _C^T}&{\theta _{{C_{\left\{ {1,2} \right\}}}}^T}&{\theta _{{C_{\left\{ {1,3} \right\}}}}^T}&{\theta _{{C_{\left\{ {2,3} \right\}}}}^T}&{\theta _{{i_1}}^T}&{\theta _{{i_2}}^T}&{\theta _{{i_3}}^T}
\end{array}} \right]^T}

\end{array}
\end{equation}
where, for $j,{j_1},{j_2} \in \left\{ {1,2,3} \right\}$ and ${j_1} \ne {j_2}$,
${K_C}\left( {\bf{P}} \right)$, ${K_{{C_{\left\{ {{j_1},{j_2}} \right\}}}}}\left( {\bf{P}} \right)$
and ${K_{{i_j}}}\left( {\bf{P}} \right)$ denote the 
sparsity of ${{\bf{z}}_C}$, ${{\bf{z}}_{{C_{\left\{ {{j_1},{j_2}} \right\}}}}}$ and ${{\bf{z}}_{{i_j}}}$ respectively. 
Likewise,
${P_C}$, 
${P_{{C_{\left\{ {{j_1},{j_2}} \right\}}}}}$ and
${P_{{i_j}}}$ are 
location matrices and
${\theta _C}$, 
${\theta _{{C_{\left\{ {{j_1},{j_2}} \right\}}}}}$ and
${\theta _{{i_j}}}$ are value vectors
of 
${{\bf{z}}_C}$, ${{\bf{z}}_{{C_{\left\{ {{j_1},{j_2}} \right\}}}}}$ and ${{\bf{z}}_{{i_j}}}$ respectively. 
From hence, we use ${\bf{P}}$ and $\Theta $ universally, not to be restricted to a specific signal model.

%
%

\section{Theoretical Bound on the Required Number of Measurements}


In this section, we find the condition on the number of measurements to recover the original signal ensembles ${\bf{X}}$ in a noiseless environment.
First, we summarize the theoretical bound of the existing DCS \cite{Baron:2009vd}
and then obtain the theoretical bound for the simple three sensors network described in Fig. \ref{fig1} in the view of the proposed GDCS model, 
which is followed by extension to the general case.


If ${\left\| {\bf{x}} \right\|_0} = K$,
the minimum number of measurements to recover the signal $\bf{x}$ is $2K$ \cite{Candes:2006eq}.
If we know the supports of the elements, it is obvious that $K$ measurements
would be sufficient for perfect recovery. 
Therefore, thinking naively, the required number of measurements for recovery is 
$\sum\limits_{j \in \Lambda } {\left\| {{{\bf{x}}_j}} \right\|{_0}} $ assuming the known supports.
However, in the DCS scenario, 
because full common information is measured by every sensor,
it is possible to recover the original signal with the number of measurements 
less than 
$\sum\limits_{j \in \Lambda } {{{\left\| {{\bf{x}}_{j}} \right\|}_0}} $.
Then the remaining problem is how to allocate the measurements to sensors to prevent
from missing the information.
Obviously, when we have ${{\bf{z}}_C}\left( n \right) \ne 0$ and ${{\bf{z}}_j}\left( n \right) \ne 0$ simultaneously, 
we cannot recover both from a single measurement. However, 
${{\bf{z}}_C}\left( n \right)$ can be recovered with help of other sensors 
whose ${{\bf{z}}_C}\left( n \right)$ does not overlap with the innovation information.
In this notion, size of overlaps for a subset of signals $\Gamma  \subseteq \Lambda $ can be quantified.

\newtheorem{definition1}{Definition}

\begin{definition1} [\cite{Baron:2009vd}, Size of overlaps] \label{def1}
The overlap size for the set of signals $\Gamma  \subseteq \Lambda $, denoted as ${K_C}\left( {\Gamma ,{\bf{P}}} \right)$, is the number of indices in which  there is overlap between the common and the innovation information supports at all signals $j \in {\Gamma ^c}$ :
\begin{equation}\label{eq12}
{K_C}\left( {\Gamma ,{\bf{P}}} \right) = \left| {\left\{ {n \in \left\{ {1,...,N} \right\}\left| {{{\bf{z}}_C}\left( n \right) \ne 0\;{\rm{ and }}\;{{\bf{z}}_j}\left( n \right) \ne 0\;{\rm{ for }}\;\forall {\rm{j}} \in {\Gamma ^c}} \right.} \right\}} \right|
\end{equation}
We also define ${K_C}\left( {\Lambda ,{\bf{P}}} \right) = {K_C}\left( {\bf{P}} \right)$
and ${K_C}\left( {\emptyset ,{\bf{P}}} \right) = 0$.
\end{definition1}
Simply, ${K_C}\left( {\Gamma ,{\bf{P}}} \right)$, for $\Gamma  \subseteq \Lambda$ implies a penalty term regarding 
the cardinality of indices
of common information
which 
should be recovered with help of measurements in $\Gamma $ due to overlaps between
common and innovation information at ${\Gamma ^c}$.
With the above definition, the theoretical required number of measurements for recovering the original signal can be determined from the following theorem.

\newtheorem{theorem1}{Theorem}

\begin{theorem1}[\cite{Baron:2009vd}, Achievable, known ${\bf{P}}$] \label{thm1}
Assume that a signal ensemble ${\bf{X}}$ is obtained from a common/innovation information JSM (Joint Sparsity Model). Let 
$M = \left( {{M_1},\;\;{M_2},\;\;...,\;{M_J}} \right)$ be a measurement tuple, and ${\left\{ {{{\bf{\Phi }}_j}} \right\}_{j \in \Lambda }}$  be random matrices having ${M_j}$ rows of i.i.d. Gaussian entries for each $j \in \Lambda$. Suppose there exists a full rank location matrix ${\bf{P}} \in {{\bf{P}}_F}\left( X \right)$ where ${{\bf{P}}_F}\left( X \right)$ is the set of feasible location matrices such that
\begin{equation} \label{eq13}
\sum\limits_{j \in \Gamma } {{M_j}}  \ge {K_{cond}}\left( {\Gamma ,{\bf{P}}} \right) = \left( {\sum\limits_{j \in \Gamma } {{K_j}\left( {\bf{P}} \right)} } \right) + {K_C}\left( {\Gamma ,{\bf{P}}} \right)
\end{equation}
for all $\Gamma  \subseteq \Lambda$. Then with probability one over ${\left\{ {{{\bf{\Phi }}_j}} \right\}_{j \in \Gamma }}$, there exists a unique solution $\hat \Theta$ to the system of equations ${\bf{Y}} = {\bf{\Phi }}{\bf{P}}\hat \Theta$; hence, the signal ensemble ${\bf{X}}$ can be uniquely recovered as ${\bf{X}} = {\bf{P}}\hat \Theta$.
\end{theorem1}

\subsection{The three sensors network using the proposed GDCS model}

The theoretical bound in the proposed GDCS model can be computed
in a similar way to Theorem \ref{thm1}. We find the required number of measurements in a subset 
$\Gamma  \subseteq \Lambda $.
A difference between the existing DCS model and the proposed GDCS model is that
there would be various types of overlaps among the signals in GDCS 
since we consider partial common information between them.

\begin{figure} [t]
\centering
\includegraphics [width=4.3in,height=3.3in]  {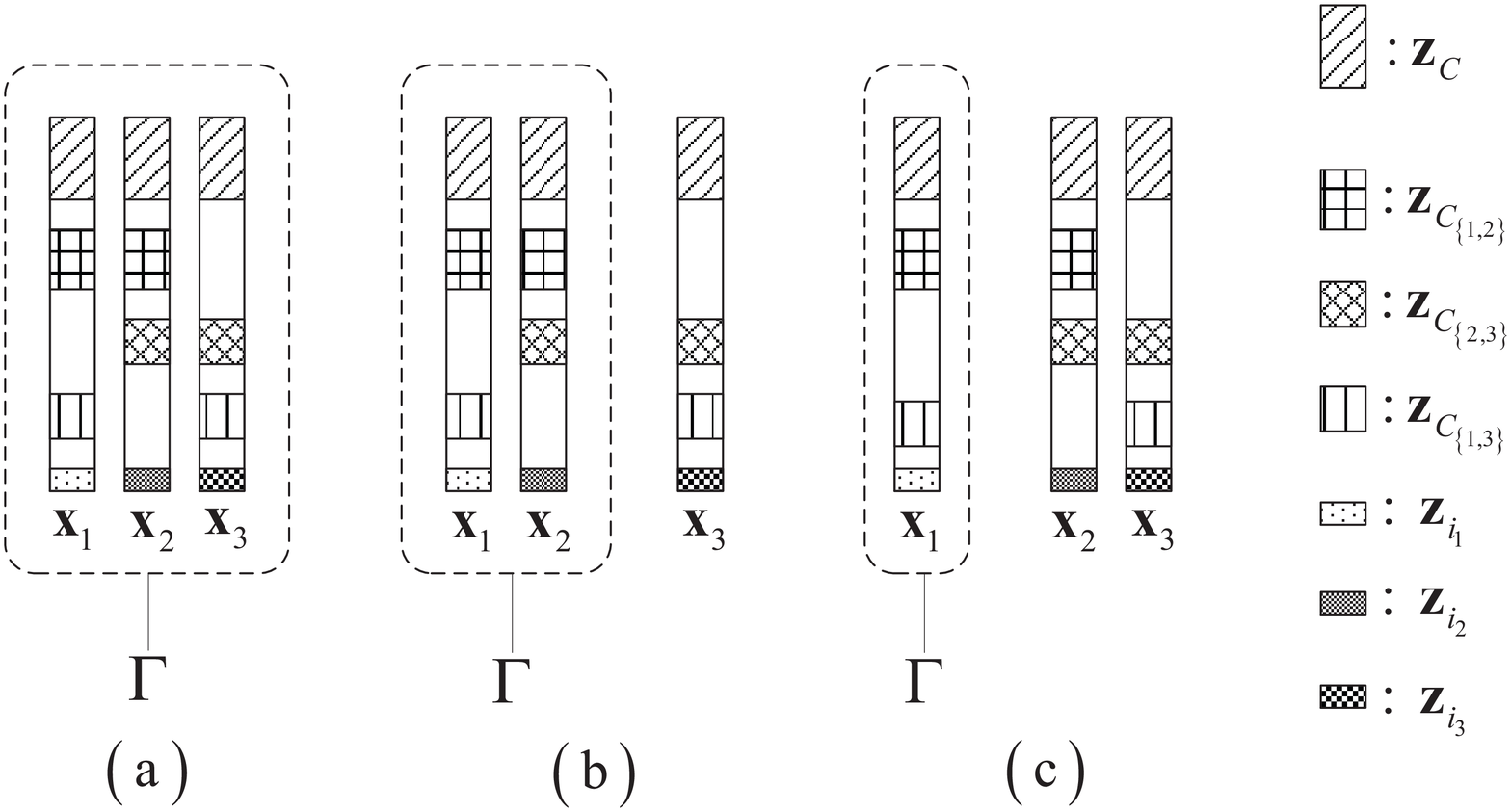}
\caption{Three possible cases of a subset $\Gamma$ including sensors.} \label{fig2}
\end{figure}

Before we go into further detail of the bound, we define the notation of partial common information.
We denote ${{\bf{z}}_{{C_\Pi }}}$ as partial common information observed by a
set of sensors $\Pi $ of cardinality $1<\left| \Pi  \right|<J$. 
For example, if partial common information is measured by a sensor set
$\left\{ {{j_1},{j_2},...,{j_\lambda }} \right\}$, 
where $\lambda $ is an arbitrary number less than $J$,
it can be represented as ${{\bf{z}}_{{C_\Pi }}}$ where 
$\Pi = \left\{ {{j_1},{j_2},...,{j_\lambda }} \right\}$.
For the three sensors network considered in Fig. \ref{fig2}, 
all the partial common information can be written
as ${{\bf{z}}_{{C_{\Pi  = \left\{ {{j_1},{j_2}} \right\}}}}}$ where 
${j_1},{j_2} \in \left\{ {1,2,3} \right\}$ and 
${j_1} \ne {j_2}$. 
Now we define two groups of information for explaining the theoretical bound.
We divide all existing information into two groups,
where the existing information includes full common, partial common and innovation information.


\begin{definition1} [Exclusive information group] \label{def2}
If the set of all sensors measuring given information is a subset of $\Gamma $,
where $\Gamma  \subseteq \Lambda $,
such information is categorized into ${\Omega _1}\left( \Gamma  \right)$.
We can write this as follows. 
\begin{equation} \label{eq14}
{\Omega _1}\left( \Gamma  \right) = \left[ {\begin{array}{*{20}{l}}
{\left\{ {{{\bf{z}}_{{i_j}}}\left| {j \cap \Gamma } \right. = j,\;j \in \Lambda } \right\}}\\
{ \cup \left\{ {{{\bf{z}}_{{C_\Pi }}}\left| {\Pi  \cap \Gamma } \right. = \Pi } \right\}}\\
{ \cup \left\{ {{{\bf{z}}_C}\left| {\Lambda  \cap \Gamma  = \Lambda } \right.} \right\}}
\end{array}} \right]
\end{equation}
\end{definition1}
We call the defined group an exclusive information group since 
the information included in this group only can be measured from the sensors belonging to $\Gamma $.
This concept can be clarified using Fig. \ref{fig2},
where each type information is symbolized, and three possible cases of a subset $\Gamma $
are shown.
In Fig. \ref{fig2}-(a), full common information 
${{\bf{z}}_C}$, partial common information ${{\bf{z}}_{{C_{\left\{ {1,2} \right\}}}}},\;{{\bf{z}}_{{C_{\left\{ {2,3} \right\}}}}},\;{{\bf{z}}_{{C_{\left\{ {1,3} \right\}}}}}$, and
innovation information ${{\bf{z}}_{{i_1}}},\;{{\bf{z}}_{{i_2}}},\;{{\bf{z}}_{{i_3}}}$ are all included in ${\Omega _1}\left( \Gamma  \right)$.
In Fig. \ref{fig2}-(b), partial common information ${{\bf{z}}_{{C_{\left\{ {1,2} \right\}}}}}$, 
and innovation information ${{\bf{z}}_{{i_1}}}$ and ${{\bf{z}}_{{i_2}}}$ are included in ${\Omega _1}\left( \Gamma  \right)$.
In Fig. \ref{fig2}-(c), only innovation information ${{\bf{z}}_{{i_1}}}$ is included in ${\Omega _1}\left( \Gamma  \right)$. 

On the contrary to this, we can define another group as follows.
\begin{definition1} [Shared information group] \label{def3}
If the set of all sensors measuring given information 
has a nonempty intersection set with $\Gamma $, where $\Gamma  \subseteq \Lambda $
but is not a subset of $\Gamma $, 
such information is categorized into ${\Omega _2}\left( \Gamma  \right)$.
We can write this as follows. 
\begin{equation} \label{eq15}
{\Omega _2}\left( \Gamma  \right) = \left[ {\begin{array}{*{20}{l}}
{\left\{ {{{\bf{z}}_{{i_j}}}\left| {j \cap \Gamma  \ne \emptyset ,\;{\kern 1pt} j \cap \Gamma  \ne j,\;j \in \Lambda } \right.} \right\}}\\
{ \cup \left\{ {{{\bf{z}}_{{C_\Pi }}}\left| {\Pi  \cap \Gamma  \ne \emptyset ,\;{\kern 1pt} \Pi  \cap \Gamma  \ne \Pi } \right.} \right\}}\\
{ \cup \left\{ {{{\bf{z}}_C}\left| {\Lambda  \cap \Gamma  \ne \emptyset ,\;{\kern 1pt} \Lambda  \cap \Gamma  \ne \Lambda } \right.} \right\}}
\end{array}} \right]
\end{equation}
\end{definition1}
We call this group a shared information group since
the information included in this group can be measured from the sensors both belonging to $\Gamma $ and not belonging to $\Gamma $.
In Fig. \ref{fig2}-(a), none of the information is included in ${\Omega _2}\left( \Gamma  \right)$.
In Fig. \ref{fig2}-(b), full common information ${{\bf{z}}_C}$ and
partial common information ${{\bf{z}}_{{C_{\left\{ {2,3} \right\}}}}},{{\bf{z}}_{{C_{\left\{ {1,3} \right\}}}}}$ are included in ${\Omega _2}\left( \Gamma  \right)$.
In Fig. \ref{fig2}-(c), full common information ${{\bf{z}}_C}$, partial common information 
${{\bf{z}}_{{C_{\left\{ {1,2} \right\}}}}}$, and ${{\bf{z}}_{{C_{\left\{ {1,3} \right\}}}}}$ are included in ${\Omega _2}\left( \Gamma  \right)$.
Lastly, we define the third group as follows.

\begin{definition1} [Unrelated information group] \label{def4}
If the set of all sensors measuring given information 
has an empty intersection set with $\Gamma $, where $\Gamma  \subseteq \Lambda $, 
such information is categorized into ${\Omega _3}\left( \Gamma  \right)$.
We can write this as follows. 
\begin{equation} \label{eq16}
{\Omega _3}\left( \Gamma  \right) = \left[ {\begin{array}{*{20}{l}}
{\left\{ {{{\bf{z}}_{{i_j}}}\left| {j \cap \Gamma } \right. = \emptyset ,\;j \in \Lambda } \right\}}\\
{ \cup \left\{ {{{\bf{z}}_{{C_\Pi }}}\left| {\Pi  \cap \Gamma } \right. = \emptyset } \right\}}\\
{ \cup \left\{ {{{\bf{z}}_C}\left| {\Lambda  \cap \Gamma  = \emptyset } \right.} \right\}}
\end{array}} \right]
\end{equation}
\end{definition1}
Since ${\Omega _3}\left( \Gamma  \right)$ is not used in obtaining the theoretical bound,
the third group has no practical meaning. Defined three groups are disjoint.

For ${\Omega _1}\left( \Gamma  \right)$, since the information included in ${\Omega _1}\left( \Gamma  \right)$ 
can be recovered only from the measurements of the sensors belonging to $\Gamma $, 
we must have the measurements of the sensors belonging to $\Gamma $ as many as sparsity of the
information included in ${\Omega _1}\left( \Gamma  \right)$.
On the other hand, for ${\Omega _2}\left( \Gamma  \right)$,
if there is no overlap,
we do not need to have the measurements of the sensors belonging to $\Gamma $
since the information included in ${\Omega _2}\left( \Gamma  \right)$
can be recovered from the measurements of the sensors not belonging to $\Gamma $. 
However, if there is an overlap, 
the information cannot be recovered from the measurements of the sensors not belonging to $\Gamma $,
so the measurements of the sensors belonging to $\Gamma $ are needed.
Therefore, we need additional measurements of the sensors belonging to $\Gamma $
to compensate these overlaps. 

Now, we explain the concept of overlap in more detail using Fig. \ref{fig2}.
Assume that we want to find the number of measurements required 
in a subset $\Gamma  \subseteq \Lambda$ for recovery. 
We assume that $\Gamma  = \Lambda $ as in Fig. \ref{fig2}-(a).
In this case, 
${{\bf{z}}_C}$, ${{\bf{z}}_{{C_{\left\{ {1,2} \right\}}}}}$, ${{\bf{z}}_{{C_{\left\{ {2,3} \right\}}}}}$, ${{\bf{z}}_{{C_{\left\{ {1,3} \right\}}}}}$, 
${{\bf{z}}_{{i_1}}}$, ${{\bf{z}}_{{i_2}}}$ and ${{\bf{z}}_{{i_3}}}$
are all included in ${\Omega _1}\left( \Gamma  \right)$, and
${\Omega _2}\left( \Gamma  \right)$ is empty.
Therefore, as mentioned above, the required number of measurements is as follows.
\begin{equation} \label{eq17}
\begin{array}{l}
\sum\limits_{j \in \Gamma } {{M_j}}  \ge {K_C}\left( {\bf{P}} \right) + \sum\limits_{{j_1},{j_2},{j_1} \ne {j_2}} {{K_{{C_{\left\{ {{j_1},{j_2}} \right\}}}}}\left( {\bf{P}} \right)}  + \sum\limits_j {{K_{{i_j}}}\left( {\bf{P}} \right)} \\
{\rm{where }}\Gamma  = \Lambda 
\end{array}
\end{equation}
The right side of inequality in (\ref{eq17}) is the sum of the sparsity of the information included in ${\Omega _1}\left( \Gamma  \right)$.
Next, let us consider the case of Fig. \ref{fig2}-(b). In this case, 
 ${{\bf{z}}_{{C_{\left\{ {1,2} \right\}}}}}$, ${{\bf{z}}_{{i_1}}}$, and ${{\bf{z}}_{{i_2}}}$ are included in ${\Omega _1}\left( \Gamma  \right)$, and
${{\bf{z}}_C}$, ${{\bf{z}}_{{C_{\left\{ {2,3} \right\}}}}}$, and ${{\bf{z}}_{{C_{\left\{ {1,3} \right\}}}}}$ are included in ${\Omega _2}\left( \Gamma  \right)$.
If there is no overlap on the information included in ${\Omega _2}\left( \Gamma  \right)$,
we need the measurements of the sensors belonging to $\Gamma$ for ${{\bf{z}}_{{C_{\left\{ {1,2} \right\}}}}}$, 
${{\bf{z}}_{{i_1}}}$, and ${{\bf{z}}_{{i_2}}}$
since other information can be recovered from the measurements of the sensors not belonging to $\Gamma $.
Therefore, the required measurements are as follows.

\begin{equation} \label{eq18}
\begin{array}{*{20}{l}}
{\sum\limits_{j \in \Gamma } {{M_j}}  \ge {K_{{C_{\left\{ {1,2} \right\}}}}}\left( {\bf{P}} \right)}
+ \sum\limits_{j = 1}^2 {{K_{{i_j}}}\left( {\bf{P}} \right)} \\
{{\rm{where}}\;\Gamma  = \left\{ {1,2} \right\}}
\end{array}
\end{equation}
The right side of the inequality in (\ref{eq18}) is the sum of the sparsity of the information included in ${\Omega _1}\left( \Gamma  \right)$.
However, assuming that there are overlaps on
the information included in ${\Omega _2}\left( \Gamma  \right)$, e.g., ${{\bf{z}}_{{C_{\left\{ {2,3} \right\}}}}}\left( n \right) \ne 0$ and ${{\bf{z}}_{{C_{\left\{ {1,3} \right\}}}}}\left( n \right) \ne 0$ for some arbitrary $n$,
more measurements than (\ref{eq18}) are needed
since the overlapped information has to be recovered with the help of the measurements 
of the sensors belonging to $\Gamma$. 
The necessary number of measurements is as follows.

\begin{equation} \label{eq19}
\begin{array}{*{20}{l}}
{\sum\limits_{j \in \Gamma } {{M_j}}  \ge {K_{{C_{\left\{ {1,2} \right\}}}}}\left( {\bf{P}} \right) + 
\sum\limits_{j = 1}^2 {{K_{{i_j}}}\left( {\bf{P}} \right)}
+Overlaps}\\
{{\rm{where}}\;\Gamma  = \left\{ {1,2} \right\}}
\end{array}
\end{equation}
where $Overlaps$ denotes additionally required number of measurements.
In the case of (c) in Fig. \ref{fig2}, similar to (\ref{eq19}), the necessary number of measurements is as follows.

\begin{equation} \label{eq20}
\begin{array}{l}
\sum\limits_{j \in \Gamma } {{M_j}}  \ge {K_{{i_1}}}\left( {\bf{P}} \right) + Overlaps\\
{\rm{where}}\;\Gamma  = \left\{ 1 \right\}
\end{array}
\end{equation}
where $Overlaps$ denotes additionally required number of measurements.
In summary, 
in order to recover the original signal perfectly, 
we need measurements of the sensors belonging to $\Gamma $ 
as many as the sparsity of the information included in ${\Omega _1}\left( \Gamma  \right)$
plus the size of overlaps of the information included in ${\Omega _2}\left( \Gamma  \right)$.

To calculate the theoretical bound on the required number of measurements analytically, we need to define the size of overlaps. 
Two types of overlaps can be differently considered: the overlaps of full common information and the overlaps of partial common information. We define the size of each type of overlap as follows.


\newtheorem{definition2}[definition1]{Definition}
\begin{definition2} [Size of overlaps of full common information] \label{def5}
Overlap size of full common information for the set of signals $\Gamma  \subseteq \Lambda$, denoted as ${O_C}\left( {\Gamma ,{\bf{P}}} \right)$, is the number of indices in which there are overlaps of the full common information and other information 
supports at all signals 
$j \in {\Gamma ^c}$. 
\begin{equation} \label{eq21}
\begin{array}{*{20}{l}}
{{O_C}\left( {\Gamma ,{\bf{P}}} \right) = ...}\\
{\left| {\begin{array}{*{20}{l}}
{\left\{ {n \in \left\{ {1,...,N} \right\}\left| {{{\bf{z}}_C}\left( n \right) \ne 0\;\;{\rm{and}}\;\;{{\bf{z}}_{{i_j}}}\left( n \right) \ne 0,\;\forall j \in {\Gamma ^c}} \right.} \right\}}\\
{ \cup \left\{ {n \in \left\{ {1,...,N} \right\}\left| {{{\bf{z}}_C}\left( n \right) \ne 0\;\;{\rm{and}}\;\;{{\bf{z}}_{{C_{\left\{ {{j_1},{j_2}} \right\}}}}}\left( n \right) \ne 0} \right.\;{\rm{for}}\;{j_1}
,{j_2}\;{\rm{such\;that}}\;{\Gamma ^c} \subseteq \left\{ {{j_{1,}}{j_2}} \right\}} \right\}}
\end{array}} \right|}
\end{array}
\end{equation}
We also define ${O_C}\left( {\Lambda ,{\bf{P}}} \right) = {K_C}\left( {\bf{P}} \right)$ and 
${O_C}\left( {\emptyset ,{\bf{P}}} \right) = 0$.
\end{definition2}
Next, we need to quantify the overlaps of partial common information.
Using the same principle as in Definition \ref{def2}, we can define size of overlaps of partial common information as follows.

\begin{definition2} [Size of overlaps of partial common information] \label{def6}
Assume that ${j_1} \in \Gamma$, and ${j_2} \in \Gamma^c$. For the set of signals $\Gamma  \subseteq \Lambda$, 
overlap size of partial common information measured by the signals $\left\{ {{j_1},{j_2}} \right\}$ i.e. 
${{\bf{z}}_{{C_{\left\{ {{j_1},{j_2}} \right\}}}}}$
, denoted as ${O_{{C_{\left\{ {{j_1},{j_2}} \right\}}}}}\left( {\Gamma ,{\bf{P}}} \right)$, 
is the number of indices for which there is a overlap between the partial common and the other information supports at a signal $j = {j_2}$. 

\begin{equation} \label{eq22}
\begin{array}{*{20}{l}}
{{O_{{C_{\left\{ {{j_1},{j_2}} \right\}}}}}\left( {\Gamma ,{\bf{P}}} \right) = ...}\\
{\left| {\begin{array}{*{20}{l}}
{\left\{ {n \in \left\{ {1,...,N} \right\}\left| {{{\bf{z}}_{{C_{\left\{ {{j_1},{j_2}} \right\}}}}}\left( n \right) \ne 0\;\;{\rm{and}}\;\;{{\bf{z}}_{{i_{j_2}}}}\left( n \right) \ne 0} \right.} \right\}}\\
{ \cup \left\{ {n \in \left\{ {1,...,N} \right\}\left| {{{\bf{z}}_{{C_{\left\{ {{j_1},{j_2}} \right\}}}}}\left( n \right) \ne 0\;\;{\rm{and}}\;\;{{\bf{z}}_{{C_{\left\{ {{j_2},{j_3}} \right\}}}}}\left( n \right) \ne 0} \right.} 
\;{\rm{for}}\;{j_3} \ne {j_1},\;{j_3} \ne {j_2}\right\}}
\end{array}} \right|}\\
{{\rm{where}}\;\;{j_1} \in \Gamma ,\;{j_2} \notin \Gamma }
\end{array}
\end{equation}
We also define ${O_{{C_{\left\{ {{j_1},{j_2}} \right\}}}}}\left( {\Lambda ,{\bf{P}}} \right) = {K_{{C_{\left\{ {{j_1},{j_2}} \right\}}}}}\left( {\bf{P}} \right)$ and
${O_{{C_{\left\{ {{j_1},{j_2}} \right\}}}}}\left( {\emptyset ,{\bf{P}}} \right) = 0$.
\end{definition2}
The reason that we only consider the overlap of partial common information which satisfy 
${j_1} \in \Gamma $ and ${j_2} \in {\Gamma ^c}$ is to consider the size of overlaps for 
partial common information included in ${\Omega _2}\left( \Gamma  \right)$.

With these definitions, we can decide the theoretical bound on the number of measurements for the three sensors example in the proposed GDCS model.

\newtheorem{theorem2}[theorem1]{Theorem}
\begin{theorem2}[Achievable, known ${\bf{P}}$] \label{thm2}
Assume that a three signal ensemble ${\bf{X}}$ is obtained from a full common/partial common/innovation information JSM (Joint Sparsity Model), as described in Fig. \ref{fig1}. Let 
$M = \left( {{M_1},\;{M_2},\;{M_3}} \right)$ be a measurement tuple, and let ${\left\{ {{{\bf{\Phi }}_j}} \right\}_{j \in \Lambda }}$  be random matrices having ${M_j}$ rows of i.i.d. Gaussian entries for each $j \in \Lambda$. Suppose there exists a full rank location matrix ${\bf{P}} \in {{\bf{P}}_F}\left( X \right)$ where ${{\bf{P}}_F}\left( X \right)$ is the set of feasible location matrices such that
\begin{equation} \label{eq23}
\begin{array}{*{20}{l}}
{\begin{array}{*{20}{l}}
{\sum\limits_{j \in \Gamma } {{M_j}}  \ge {O_C}\left( {\Gamma ,{\bf{P}}} \right) + \sum\limits_{{j_1} \in \Gamma ,{j_2} \in {\Gamma ^c},{j_1} \ne {j_2}} {{O_{{C_{\left\{ {{j_1},{j_2}} \right\}}}}}\left( {\Gamma ,{\bf{P}}} \right)}  + \;...}\\
{\;\;\;\;\;\;\;\;\;\;\;\sum\limits_{{j_1} \in \Gamma ,{j_2} \in \Gamma ,{j_1} \ne {j_2}} {{K_{{C_{\left\{ {{j_1},{j_2}} \right\}}}}}\left( {\bf{P}} \right)}  + \sum\limits_{j \in \Gamma }^{} {{K_{{i_j}}}\left( {\bf{P}} \right)} }
\end{array}}\\
{\;\;\;\;\;}
\end{array}
\end{equation}
for all $\Gamma  \subseteq \Lambda$. Then, with probability one, there exists a unique solution $\hat \Theta$ to the system of equations ${\bf{Y}} = {\bf{\Phi }}{\bf{P}}\hat \Theta$ ; hence, the signal ensemble ${\bf{X}}$ can be uniquely recovered as ${\bf{X}} = {\bf{P}}\hat \Theta$.  
\end{theorem2}

Assuming a subset $\Gamma  \subseteq \Lambda $, 
arbitrary partial common information ${{\bf{z}}_{{C_{\left\{ {{j_1},{j_2}} \right\}}}}}$ is included in ${\Omega _1}\left( \Gamma  \right)$ 
if  ${j_1} \in \Gamma ,\;{j_2} \in \Gamma $, and it is included in ${\Omega _2}\left( \Gamma  \right)$
if ${j_1} \in \Gamma ,\;{j_2} \in {\Gamma ^c}$. Therefore, for the partial common information ${{\bf{z}}_{{C_{\left\{ {{j_1},{j_2}} \right\}}}}}$
which satisfies ${j_1} \in \Gamma ,\;{j_2} \in \Gamma $, we consider the sparsity as the required number of measurements,
and for the partial common information ${{\bf{z}}_{{C_{\left\{ {{j_1},{j_2}} \right\}}}}}$
which satisfies ${j_1} \in \Gamma ,\;{j_2} \in \Gamma^c $, we consider the size of overlaps as the required number of measurements.

\subsection{The general case}

The case of a larger number of sensors can be readily extended from the three sensors networks. 
With a larger number of sensors, many various kinds of partial common information can be characterized 
depending on how they share the information.
%
Now, we generalize the size of overlaps to derive the bound on the number of
measurements.

\begin{definition2} [Size of overlaps of full common information, the general version] \label{def7}
The overlap size of full common information,
 denoted as ${O_C}\left( {\Gamma ,{\bf{P}}} \right)$, 
is the number of indices in which there is overlap between the full common and other information supports at all signals $j \in \Gamma^c$.
\begin{equation} \label{eq24}
\begin{array}{*{20}{l}}
{\begin{array}{*{20}{l}}
{{O_C}\left( {\Gamma ,{\bf{P}}} \right) = ...}\\
{\left| {\begin{array}{*{20}{l}}
{\left\{ {n \in \left\{ {1,...,N} \right\}\left| {{{\bf{z}}_C}\left( n \right) \ne 0\;\;{\rm{and}}\;\;{{\bf{z}}_{{i_j}}}\left( n \right) \ne 0,\;\forall j \in {\Gamma ^c}} \right.} \right\}}\\
{ \cup \left\{ {n \in \left\{ {1,...,N} \right\}\left| {{{\bf{z}}_C}\left( n \right) \ne 0\;\;{\rm{and}}\;\;{{\bf{z}}_{{C_{{\Pi _k}}}}}\left( n \right) \ne 0,\;\forall k\;{\rm{such\;that}}\;
{\Gamma ^c} \subseteq \bigcup\limits_k {{\Pi _k}}  } \right.} \right\}}\\
{ \cup \left\{ {n \in \left\{ {1,...,N} \right\}\left| {\begin{array}{*{20}{l}}
{{{\bf{z}}_C}\left( n \right) \ne 0\;\;{\rm{and}}\;\;{{\bf{z}}_{{C_{{\Pi _{k}}}}}}\left( n \right) \ne 0}\\
{\rm{and}}\;{{{\bf{z}}_{{i_{j}}}}\left( n \right) \ne 0,\;\forall j \in {\Gamma ^c} -{\left( {\bigcup\limits_k {{\Pi _k}}} \right)}}
\end{array}} \right.} \right\}}
\end{array}} \right|}
\end{array}}
\end{array}
\end{equation}
where ${\Pi _k}$ is a sensor set for partial common information ${{\bf{z}}_{{C_{{\Pi _k}}}}}$ and
$\bigcup\limits_k {{\Pi _k}}$ is a union of sensor sets for partial common information. 
We also define ${O_C}\left( {\Lambda ,{\bf{P}}} \right) = {K_C}\left( {\bf{P}} \right)$
and ${O_C}\left( {\emptyset ,{\bf{P}}} \right) = 0$.
\end{definition2}
(\ref{eq24}) consists of a union of three sets. 
The sets present overlaps between 
the full common information and the innovation information, overlaps between the full common information and the partial common information,
and overlaps between the full common information and both of the innovation information and the partial common information, respectively.
Now, Definition \ref{def6} should be extended to the general case.

\begin{definition2} [Size of overlaps of partial common information, the general version] \label{def8}
The overlap size of partial common information measured by a sensor set $\Pi $, i.e., ${{\bf{z}}_{{C_\Pi }}}$,
 denoted as ${O_{{C_\Pi }}}\left( {\Gamma ,{\bf{P}}} \right)$, is the number of indices for which there is overlap between the partial common and other information supports at all signals $j \in \left( {\Pi  \cap {\Gamma ^c}} \right)$. 
\begin{equation} \label{eq25}
\begin{array}{l}
{O_{{C_\Pi }}}\left( {\Gamma ,{\bf{P}}} \right) = ...\\
\left| \begin{array}{l}
\left\{ {n \in \left\{ {1,...,N} \right\}\left| {{{\bf{z}}_{{C_\Pi }}}\left( n \right) \ne 0\;{\rm{ and }}\;{{\bf{z}}_{{i_j}}}\left( n \right) \ne 0,{\rm{ }}\forall j \in \left( {\Pi  \cap {\Gamma ^c}} \right)} \right.} \right\}\\
 \cup \left\{ {n \in \left\{ {1,...,N} \right\}\left| {{{\bf{z}}_{{C_\Pi }}}\left( n \right) \ne 0\;{\rm{ and }}\;{{\bf{z}}_{{C_{{\Pi _k}}}}}\left( n \right) \ne 0,{\rm{ }}\forall k\;{\rm{ such \;that }}\;\left( {\Pi  \cap {\Gamma ^c}} \right) \subseteq \bigcup\limits_k {{\Pi _k}} } \right.{\rm{ }}} \right\}\\
\left\{ {n \in \left\{ {1,...,N} \right\}\left| \begin{array}{l}
{{\bf{z}}_{{C_\Pi }}}\left( n \right) \ne 0\;{\rm{ and }}\;{{\bf{z}}_{{C_{{\Pi _k}}}}}\left( n \right) \ne 0\\
{\rm{and }}\;{{\bf{z}}_{{i_j}}}\left( n \right) \ne 0,{\rm{ }}\forall j \in \left( {\Pi  \cap {\Gamma ^c}} \right) - \left( {\bigcup\limits_k {{\Pi _k}} } \right)
\end{array} \right.} \right\}
\end{array} \right|
\end{array}
\end{equation}
where ${\Pi _k}$ is a sensor set for partial common information ${{\bf{z}}_{{C_{{\Pi _k}}}}}$ and
$\bigcup\limits_k {{\Pi _k}} $ is a union of the sensor sets for partial common information. 
We also define ${O_{{C_\Pi }}}\left( {\Lambda ,{\bf{P}}} \right) = {K_{{C_\Pi }}}\left( {\bf{P}} \right)$ and
${O_{{C_\Pi}}}\left( {\emptyset ,{\bf{P}}} \right) = 0$.
\end{definition2}
As in (\ref{eq24}), (\ref{eq25}) consists of union of three sets, and each set is a case of overlaps.
The first set is overlaps between the partial common information and the innovation information, 
the second set is overlaps between the partial common information and other partial common information, and
the third set is overlaps between the partial common information and both of the innovation information and other partial common information.

With these definitions, we can compute the theoretical bound on 
the required number of measurements for the proposed GDCS model for the general case.
\begin{theorem2}[Achievable, known ${\bf{P}}$] \label{thm3}
Assume that a signal ensemble ${\bf{X}}$ is obtained from a full common/partial common/innovation information JSM (Joint Sparsity Model).
Let $M = \left\{ {{M_1},\;{M_2},\;...,\;{M_J}} \right\}$ be a measurement tuple, and let ${\left\{ {{{\bf{\Phi }}_j}} \right\}_{j \in \Lambda }}$  be random matrices having ${M_j}$ rows of i.i.d. Gaussian entries for each $j \in \Lambda$. Suppose there exists a full rank location matrix ${\bf{P}} \in {{\bf{P}}_F}\left( X \right)$ where ${{\bf{P}}_F}\left( X \right)$ is the set of feasible location matrices such that
\begin{equation} \label{eq26}
\begin{array}{*{20}{l}}
{\sum\limits_{j \in \Gamma } {{M_j} \ge {O_C}\left( {\Gamma ,{\bf{P}}} \right) + \sum\limits_{{{\bf{z}}_{{C_\Pi }}} \in \;{\Omega _2}\left( \Gamma  \right)} {{O_{{C_\Pi }}}\left( {\Gamma ,{\bf{P}}} \right)}  + \;\;.} ..}\\
{\;\;\;\;\;\;\;\;\;\;\;\sum\limits_{{{\bf{z}}_{{C_\Pi }}} \in \;{\Omega _1}\left( \Gamma  \right)} {{K_{{C_\Pi }}}\left( {\bf{P}} \right)}  + \sum\limits_{j \in \Gamma } {{K_{{i_j}}}\left( {\bf{P}} \right)} }
\end{array}
\end{equation}
for all $\Gamma  \subseteq \Lambda$. Then, with probability one, there exists a unique solution $\hat \Theta$ to the system of equations ${\bf{Y}} = {\bf{\Phi }}{\bf{P}}\hat \Theta$ ; hence, the signal ensemble ${\bf{X}}$ can be uniquely recovered as ${\bf{X}} = {\bf{P}}\hat \Theta$.  
\end{theorem2}
The proof is described in the Appendix.
As in Theorem \ref{thm2}, for the information included in ${\Omega _1}\left( \Gamma  \right)$, we consider the sparsity of the information as the required number of measurements.
On the other hand, for the information included in ${\Omega _2}\left( \Gamma  \right)$, 
we consider the size of overlaps of the information as the required number of measurements.

We can observe that Theorem \ref{thm1}, the theoretical bound of the existing DCS model,
is a special case of Theorem \ref{thm3}, i.e., the case not considering partial common information.
Therefore, Theorem \ref{thm3} can be regarded as a more refined version of Theorem \ref{thm1}.
When $\bf{P}$ is unknown, it is known that additional $\left| \Gamma  \right|$ measurements 
in the right side of (\ref{eq26}) would be sufficient for recovery \cite{Baron:2009vd}.

\section{Iterative Signal Detection With Sequential Correlation Search}

In this section, we discuss a method that can benefit from partial common information 
without any a priori-knowledge about correlation structure, which is 
the main obstacle of exploiting partial common information 
in practical implementation.
To compare the requirement of a priori-knowledge of
the existing DCS and the proposed GDCS, 
the problem formulation of the existing DCS model is described as follows.
The notations of the information follow the existing DCS style.

\begin{equation}\label{eq27}
{\bf{X}} = {\left[ {\begin{array}{*{20}{c}}
{{\bf{x}}_1^T}&{{\bf{x}}_2^T}& \cdots &{{\bf{x}}_J^T}
\end{array}} \right]^T} \in {\mathbb{R}^{NJ}}
\end{equation}
\begin{equation} \label{eq28}
{\bf{Z}}: = {\left[ {\begin{array}{*{20}{c}}
{{\bf{z}}_C^T}&{{\bf{z}}_1^T}&{{\bf{z}}_2^T}& \cdots &{{\bf{z}}_J^T}
\end{array}} \right]^T} \in {\mathbb{R}^{\left( {J + 1} \right)N}}
\end{equation}
\begin{equation} \label{eq29}
{{\bf{x}}_j} = {{\bf{z}}_C} + {{\bf{z}}_j},\;{\rm{where}}\;j \in \Lambda 
\end{equation}
\begin{equation} \label{eq30}
{\tilde \Phi : = \left[ {\begin{array}{*{20}{c}}
{{\Phi _1}}&{{\Phi _1}}&0& \cdots &0\\
{{\Phi _2}}&0&{{\Phi _2}}& \cdots &0\\
 \vdots & \vdots & \vdots & \ddots & \vdots \\
{{\Phi _J}}&0&0& \cdots &{{\Phi _J}}
\end{array}} \right] \in {\mathbb{R}^{JM \times \left( {J + 1} \right)N}}}
\end{equation}
\begin{equation} \label{eq31}
{{\bf{Y}} = \tilde \Phi {\bf{Z}}}
\end{equation}
\begin{equation} \label{eq32}
\begin{array}{l}
{\bf{\hat{Z}}} = \arg \min {\left\| {{{\bf{W}}_C}{\bf{z}}_C'} \right\|_1} + {\left\| {{{\bf{W}}_1}{\bf{z}}_1'} \right\|_1} +  \cdots  + {\left\| {{{\bf{W}}_J}{\bf{z}}_J'} \right\|_1}\\
s.t.\;{\kern 1pt} \;{\kern 1pt} {\bf{Y}} = \tilde \Phi {{\bf{Z}}'}
\end{array}
\end{equation}
where ${{\bf{W}}_C}$ and ${{\bf{W}}_j}$, $j \in \Lambda $ are weight matrices.
Thanks to a joint recovery, the improved recovery performance can be obtained compared to 
separate recovery.

To get some insight on the proposed algorithm, let us consider a case in which partial common information is measured 
by a set of sensors $\Lambda \backslash \left\{ {1,2,3} \right\}$. 
This case can be formulated as the following problem by using the proposed GDCS model.


\begin{equation} \label{eq33}
{\bf{X}} = {\left[ {\begin{array}{*{20}{c}}
{{\bf{x}}_1^T}&{{\bf{x}}_2^T}& \cdots &{{\bf{x}}_J^T}
\end{array}} \right]^T} \in {\mathbb{R}^{NJ}}
\end{equation}
\begin{equation} \label{eq34}
{\bf{Z}}: = {\left[ {\begin{array}{*{20}{c}}
{{\bf{z}}_{{C_\Pi }}^T}&{{\bf{z}}_{{i_1}}^T}&{{\bf{z}}_{{i_2}}^T}& \cdots &{{\bf{z}}_{{i_J}}^T}
\end{array}} \right]^T} \in {\mathbb{R}^{\left( {J + 1} \right)N}},\;{\rm{where}}\;\Pi  = \Lambda \backslash \left\{ {1,2,3} \right\}
\end{equation}
\begin{equation} \label{eq35}
{{\bf{x}}_j} = \left\{ {\begin{array}{*{20}{c}}
{{{\bf{z}}_{{i_j}}}\;\;{\rm{if }}\;j \notin \Pi }\\
{{{\bf{z}}_{{C_\Pi }}} + {{\bf{z}}_{{i_j}}}\;\;{\rm{if }}\;j \in \Pi }
\end{array}} \right.
\end{equation}
\begin{equation} \label{eq36}
\tilde \Phi  = \left[ {\begin{array}{*{20}{c}}
0&{{\Phi _1}}&0&0&0& \cdots &0\\
0&0&{{\Phi _2}}&0&0& \cdots &0\\
0&0&0&{{\Phi _3}}&0& \cdots &0\\
{{\Phi _4}}&0&0&0&{{\Phi _4}}& \cdots &0\\
 \vdots & \vdots & \vdots & \vdots & \vdots & \ddots &0\\
{{\Phi _{J}}}&0&0&0&0&0&{{\Phi _{J}}}
\end{array}} \right] \in {\mathbb{R}^{JM \times \left( {J + 1} \right)N}}
\end{equation}
\begin{equation} \label{eq37}
{{\bf{Y}} = \tilde \Phi {\bf{Z}}}
\end{equation}
\begin{equation} \label{eq38}
\begin{array}{l}
{\bf{\hat Z}} = \arg \min {\left\| {{{\bf{W}}_{{C_\Pi }}}{\bf{z}}{'_{{C_\Pi }}}} \right\|_1} + {\left\| {{{\bf{W}}_{{i_1}}}{\bf{z}}{'_{{i_1}}}} \right\|_1} +  \cdots  + {\left\| {{{\bf{W}}_{{i_J}}}{\bf{z}}{'_{{i_J}}}} \right\|_1}\\
s.t.\;\;{\bf{Y}} = \tilde \Phi {\bf{Z}}'
\end{array}
\end{equation}
where ${{\bf{W}}_{C_{\Pi}}}$ and ${{\bf{W}}_{i_j}}$, $j \in \Lambda $ are weight matrices.
As shown above, to exploit partial common information, we have to find a sensor set for partial common information $\Pi $, 
in this case $\Lambda \backslash \left\{ {1,2,3} \right\}$.
Unfortunately, it is not straightforward to find which sensors are correlated.
Since each sensor compresses its signal without cooperation of other sensors,
there is nothing we can do to determine the correlation structure in a compression process. In a recovery process,
although we can find the correlation structure by 
an exhaustive search, it demands approximately ${2^J}$ number of searches, which is not practical.
Therefore, we need an moderately complex algorithm that finds the correlation structure.
 
A novel algorithm is proposed for 
finding the correlation structure.
The algorithm iteratively selects the least correlated sensor so that 
we can approximate the sensor set for partial common information $\Pi$.
For simplicity, we assume a joint sparse signal ensemble $\bf{X}$ with partial common information 
${{\bf{z}}_{{C_\Pi }}}$, where $\Pi  = \Lambda \backslash \left\{ {1,2,3} \right\}$ as in (\ref{eq34}).
However, since we have no knowledge on the correlation structure, 
we cannot formulate the measurement matrix as in (\ref{eq36}).
Instead, we use the solutions of the separate recovery and of the existing DCS framework which considers only full common information.
The solution obtained by using the existing DCS has to include full common information,
although the given signal ensemble $\bf{X}$ may not have full common information.
By using this forcefully found full common information, 
we can obtain a clue about the correlation structure.
When we compare $l_1$ norm of the innovation information between the solution vector of the separate recovery and the existing DCS, 
while the $l_1$ norm tends to increase
($l_1$ norm of the innovation information of the solution vector obtained using the existing DCS becomes larger.)
if the corresponding sensor $j \notin \Pi$,
it tends to be decreased if the corresponding sensor $j \in \Pi $.

Though this phenomenon is difficult to understand at first glance, 
it is quite straightforward. 
We should note that
the forcefully found full common information 
may have some relation with the true partial common information.
Actually, the forcefully found full common information 
is likely to be similar to
the partial common information to minimize $l_1$ norm of the solution vector.
(But not always. We will explain it after this paragraph.)
Then, if the sensor $j \in \Pi$, which means it is one of the sensors that measure the partial common information,
a joint recovery process successfully divides the energy of the signal into a joint recovery part (the first column of $\tilde \Phi $ in (\ref{eq30}))
and a separate recovery part (the rest of the columns of $\tilde \Phi $ in (\ref{eq30})).
However, if the sensor $j \notin \Pi$, which means it is one of the sensors that do not have the partial common information,
the innovation information of the sensor $j \notin \Pi$ must be made to compensate 
the forcefully found full common information, causing the increase in $l_1$ norm of the innovation information.

However,  
it can be exploited only if forcefully found full common information is similar to partial common information.
If only a small number of sensors can measure partial common information, i.e., $\left| \Pi  \right|$ is small, 
the forcefully found full common information has no relationship to the partial common information. In this case,
we cannot expect to find the sensor set $\Pi $ based on the above observation.
Therefore, in this paper,
we assume that any partial common information can be measured by a sufficient number of sensors.
This assumption can be justified by the fact that significant performance gain by joint recovery
of partial common information can be achieved when a sufficient number of sensors
measure the partial common information.

Exploiting the above intuition, 
an iterative signal detection with a sequential correlation search algorithm is proposed with 
an underlying assumption on arbitrary inter-signal correlation 
(full common information or partial common information).
The algorithm assumes the following.
The number of sensors is $J$, each signal size is $N$ and arbitrary inter-signal correlation (full common information or partial common information) exists.
The received signal is denoted by ${\bf{Y}} \in {\mathbb{R}^{MJ}}$, where $M$ is the number of measurements assigned to each sensor node.

\newtheorem{algorithm1}{Algorithm}
\begin{algorithm1}\ \\
0: Set the iteration counts $i_1$ and ${i_2}$ to zero and initialize a matrix ${{\bf{I}}_{{i_2}}} = {\left[ {\begin{array}{*{20}{c}}
{\Phi _1^T}& \cdots &{\Phi _J^T}
\end{array}} \right]^T}$.
Define a matrix ${\bf{H}}\left( j \right) \in {\mathbb{R}^{MJ \times N}},\;{\kern 1pt} j \in \Lambda $ such that it holds ${\Phi _j}$ on the $j$th block position
while setting other blocks to zero.
For example,
\begin{equation} \label{eq39}
{\bf{H}}\left( j \right) = {\left[ {\begin{array}{*{20}{c}}
{{0^T}}& \cdots &{\Phi _j^T}& \cdots &{{0^T}}
\end{array}} \right]^T}
\end{equation}
Initialize a set of indexes of the sensors  ${\Sigma _{{i_2}}} = \emptyset $, 
variables $\alpha_{{i_2}}  = 0$ and ${\beta _{{i_1}}} = 0$ and 
a matrix $\tilde \Phi _{{i_1}}^{update}={\bf{0}}$.
The $\alpha_{{i_2}} $ and $\beta_{{i_1}}$ variables will be used as a parameter for escaping the loop
and $\tilde \Phi _{{i_1}}^{update}$ will be used as a temporary matrix for an updated measurement matrix.
The notation ``$ \Leftarrow $" means to substitute the left hand side parameter
with the right hand side parameter.
\\
1: Construct two measurement matrices as follows.
\begin{equation} \label{eq40}
\tilde \Phi _{{i_2}}^1 = \left[ {{{\bf{I}}_{{i_2}}}\left| {\begin{array}{*{20}{c}}
{{\Phi _1}}&0& \cdots &0\\
0&{{\Phi _2}}& \cdots &0\\
 \vdots & \vdots & \ddots & \vdots \\
0&0& \cdots &{{\Phi _J}}
\end{array}} \right.} \right],\;\;\\
\tilde \Phi^2 = \left[ {\begin{array}{*{20}{c}}
{{\Phi _1}}&0& \cdots &0\\
0&{{\Phi _2}}& \cdots &0\\
 \vdots & \vdots & \ddots & \vdots \\
0&0& \cdots &{{\Phi _J}}
\end{array}} \right]
\end{equation}\\
2: Obtain ${{\bf{\hat Z}}_1}$ and ${{\bf{\hat Z}}_2}$ by solving the following weighted $l_1$ minimization problems.
\begin{equation} \label{eq41}
\begin{array}{*{20}{l}}
{{{{\bf{\hat Z}}}_1} = \arg \min {{\left\| {{{\bf{W}}_1}{{\bf{Z}}_1}} \right\|}_1}}\\
{s.t.\;\;{\bf{Y}} = \tilde \Phi _{{i_2}}^1{{\bf{Z}}_1}}
\end{array}
\end{equation}
\begin{equation} \label{eq42}
\begin{array}{l}
{{\bf{\hat Z}}_2} = \arg \min {\left\| {{\bf{W}}_2{{{\bf{Z}}}_2}} \right\|_1}\\
s.t.\;\;{\bf{Y}} = {{\tilde \Phi }^2}{{\bf{ Z}}_2}
\end{array}
\end{equation}
where ${\bf{W}}_j$, $j = 1\;or\;2$ are weight matrices. To avoid confusion, we denote the innovation information part of each solution
in the following forms. The innovation information part indicates a part of the solution vector that is multiplied by the block diagonal part of 
${\tilde \Phi _{{i_2}}}^1$ or ${\tilde \Phi ^2}$.
\begin{equation} \label{eq43}
\begin{array}{l}
{{\bf{\hat Z}}_1} = {\left[ {\begin{array}{*{20}{c}}
 \cdots &{{{\left( {{\bf{\hat z}}_1^1} \right)}^T}}& \cdots &{{{\left( {{\bf{\hat z}}_J^1} \right)}^T}}
\end{array}} \right]^T}
,\;\;{\bf{\hat z}}_j^1 \in {\mathbb{R}^{N \times 1}},\;j \in \Lambda\\
{{\bf{\hat Z}}_2} = {\left[ {\begin{array}{*{20}{c}}
 \cdots &{{{\left( {{\bf{\hat z}}_1^2} \right)}^T}}& \cdots &{{{\left( {{\bf{\hat z}}_J^2} \right)}^T}}
\end{array}} \right]^T}
,\;\;{\bf{\hat z}}_j^2 \in {\mathbb{R}^{N \times 1}},\;j \in \Lambda
\end{array}
\end{equation}\\
3: $\alpha_{{i_2}+1}  \Leftarrow {\left\| {{{{\bf{\hat Z}}}_1}} \right\|_0} $. (We treat ${{\bf{\hat Z}}_1}\left( n \right)$ as zero if 
$\left| {{{{\bf{\hat Z}}}_1}\left( n \right)} \right| < {10^{ - 4}}$.)\\
4: Find the sensor index ${j^*}$ by solving the following problem.
\begin{equation} \label{eq44}
\begin{array}{l}
{j^*} = \arg \min {\left\| {{\bf{\hat z}}_j^2} \right\|_1} - {\left\| {{\bf{\hat z}}_j^1} \right\|_1}\\
s.t.\;\;j \in  \Lambda   - \Sigma _{{i_2}}
\end{array}
\end{equation}
We denote the obtained index as ${j_{\min }}$.\\
5: ${\bf{I}}_{{i_2} + 1}^{} \Leftarrow {{\bf{I}}_{{i_2}}} - {\bf{H}}\left( {{j_{\min }}} \right)$.\\
6: ${\tilde \Phi _{{i_2}+1}^1} \Leftarrow \left[ {{{\bf{I}}_{{i_2} + 1}}\left| {\begin{array}{*{20}{c}}
{{\Phi _1}}&0& \cdots &0\\
0&{{\Phi _2}}& \cdots &0\\
 \vdots & \vdots & \ddots & \vdots \\
0&0& \cdots &{{\Phi _J}}
\end{array}} \right.} \right]$.\\
7:  ${\Sigma _{{i_2} + 1}} \Leftarrow \left\{ {\begin{array}{*{20}{c}}
{{\Sigma _{{i_2}}}}&{{j_{\min }}}
\end{array}} \right\}$.\\
8: ${i_2} \Leftarrow {i_2} + 1$.\\
9: Go to step 2 and repeat the loop until the following two conditions are satisfied.
\begin{equation} \label{eq45}
\begin{array}{l}
1 < {i_2}\\
{\alpha _{{i_2} - 1}} < {\alpha _{{i_2} }}
\end{array}
\end{equation}
10: $\beta_{{i_1}+1} \Leftarrow \alpha_{{i_2}-1}$.\\
11: $\tilde \Phi _{{i_1}+1}^{update} \Leftarrow {\tilde \Phi _{{i_2}-1}^1}$.\\
12: ${i_1} \Leftarrow {i_1} + 1$, ${i_2} \Leftarrow 0$.\\
13: ${{\bf{I}}_{{i_2}}} \Leftarrow {\left[ {\begin{array}{*{20}{c}}
{\Phi _1^T}& \cdots &{\Phi _J^T}
\end{array}} \right]^T}$, ${\Sigma _{{i_2}}} \Leftarrow \emptyset $.\\
14. With $\tilde \Phi _{{i_1}}^{update}$, update each matrix as follows.

\begin{equation} \label{eq46}
\begin{array}{*{20}{l}}
{{{\tilde \Phi }_{{i_2}}^1} \Leftarrow \left[ {{{\bf{I}}_{{i_2}}}\left| {\tilde \Phi _{{i_1}}^{update}} \right.} \right]}\\
{{{\tilde \Phi }^2} \Leftarrow \tilde \Phi _{{i_1}}^{update}}
\end{array}
\end{equation}
15: Go to step 2 and repeat the loop until the following conditions are satisfied.
\begin{equation} \label{eq47}
\begin{array}{l}
1 < {i_1}\\
{\beta _{{i_1} - 1}} < {\beta _{{i_1} }}
\end{array}
\end{equation}
16: Estimate the signal by solving the following problem.

\begin{equation} \label{eq48}
\begin{array}{l}
{\bf{\hat Z}} = \arg \min {\left\| {{\bf{WZ}}} \right\|_1}\\
s.t.\;{\bf{Y}} = \tilde \Phi _{{i_1} - 1}^{update}{\bf{Z}}
\end{array}
\end{equation}

\end{algorithm1}

The algorithm consists of two phases, the inner and outer phases.
In the inner phase, 
starting from an assumption of full common information,
we exclude sensors one by one from the candidate sensor set, based on (\ref{eq44}).
In the outer phase, we search out different types of inter-signal correlation.

Intuitively, as the algorithm proceeds, we pursue
the measurement matrix corresponding to a more sparse solution
so that the algorithm provides better performance.
However, computational complexity linearly increases with the number of iterations.
In the worst case, the proposed algorithm needs $J$ iterations in each inner phase.
For the outer phase, since the number of columns of the measurements matrix $\tilde \Phi _{{i_2}}^1$ 
is increased by one with every iteration,
we can describe the complexity of our algorithm as $O\left( J{{{\left( {N(J + \mu )} \right)}^3}} \right)$
where $\mu$ is the number of iterations of the outer phase in the algorithm.
If the algorithm searches the correlation structure exactly,
$\mu$ would be the number of inter-signal correlations.
The existing DCS has complexity $O\left( {{{\left( {N(J + 1)} \right)}^3}} \right)$.
In the real simulation, however, 
the decoding time is necessarily limited 
since the stopping criterion is the approximated ${l_0}$ norm of the solution vector which is a natural number.
The iteration continues only if the approximated ${l_0}$ norm of the solution vector is reduced compared to the previous iteration,
which implies that the maximum iteration number is the approximated ${l_0}$ norm of the solution vector of the first iteration.
The real CPU times consumed by the algorithm are compared in the next section.

\section{Numerical Experiments}



\begin{figure}
\begin{center}$
\begin{array}{cc}
\includegraphics [width=5in,height=3.7in]  {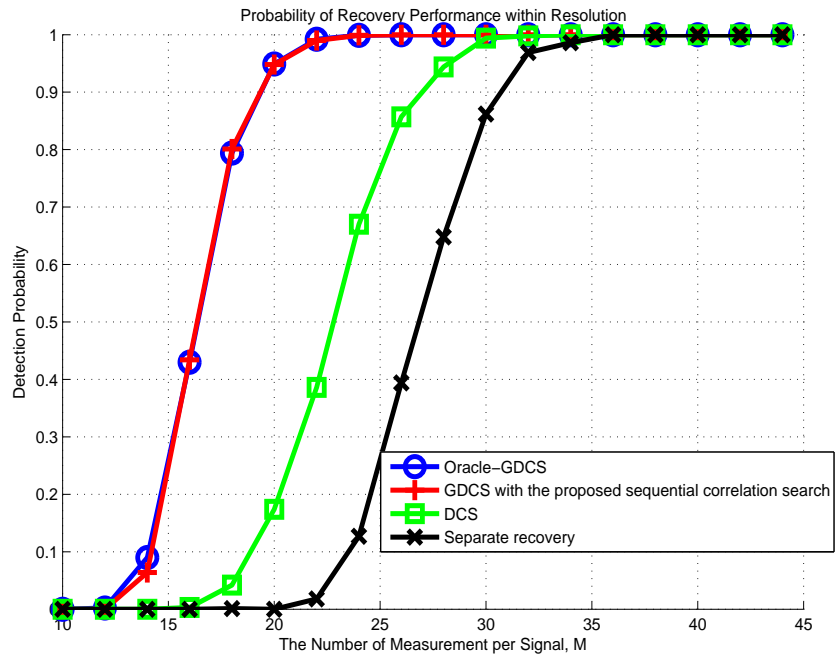} \\
\mbox{(a)}\\
\includegraphics [width=5in,height=3.7in]  {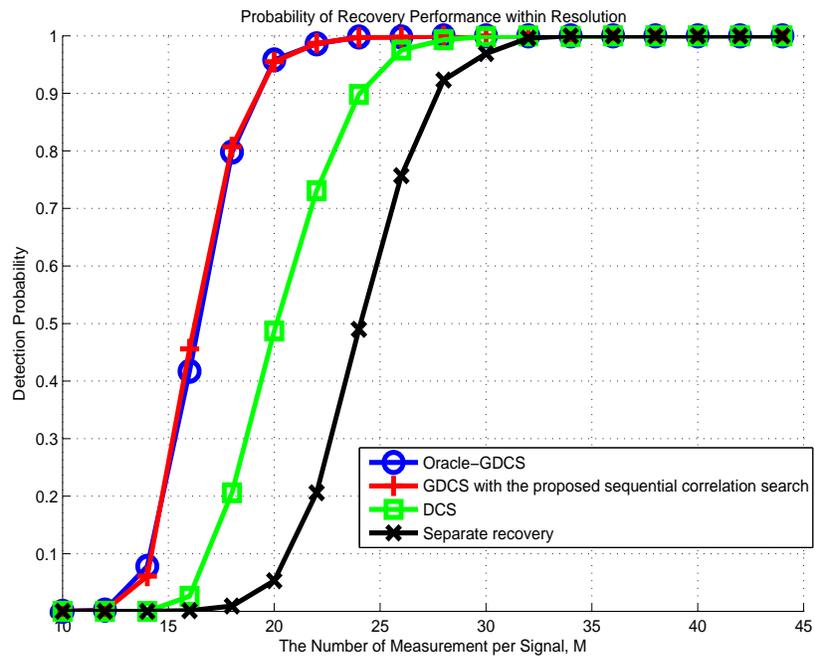} \\
\mbox{(b)}\\
\end{array}$
\end{center}
\caption{Performance comparison of Oracle GDCS, GDCS with the proposed algorithm, the existing DCS and separate recovery when (a) ${K_{{i_j}}} = 4,\;{K_{{C_\Pi }}} = 6,\;\left| \Pi  \right| = 6$, (b) ${K_{{i_j}}} = 4,\;{K_{{C_\Pi }}} = 4,\;\left| \Pi  \right| = 6$} \label{fig3}
\end{figure}

\begin{figure}
\begin{center}$
\begin{array}{cc}
\includegraphics [width=5in,height=3.7in]  {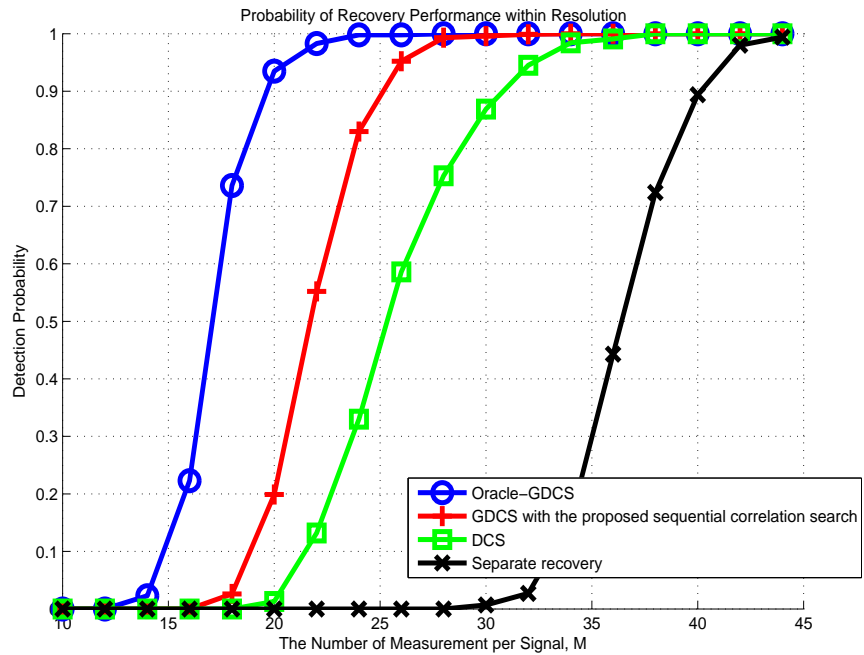}\\
\mbox{(a)} \\
\includegraphics [width=5in,height=3.7in]  {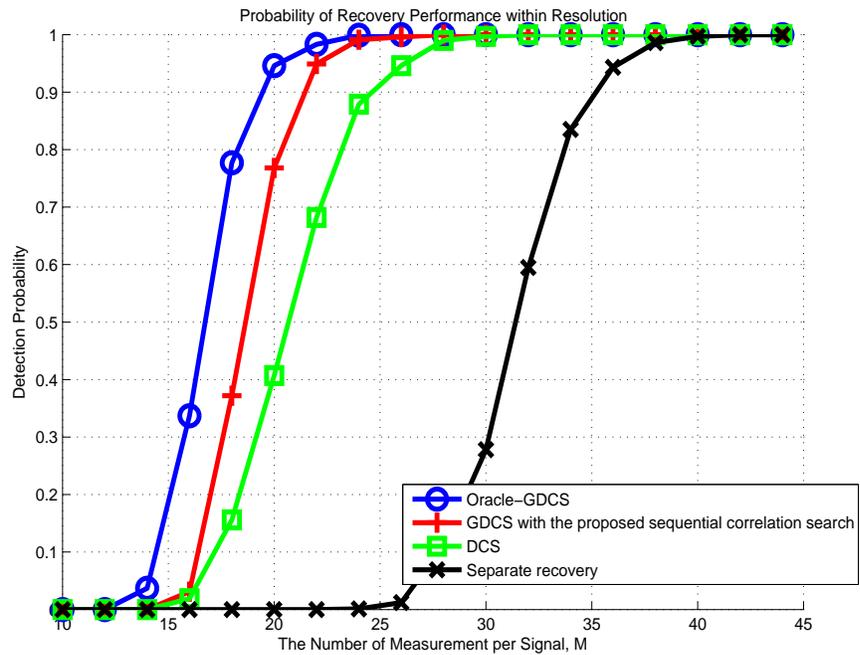} \\
\mbox{(b)}\\
\end{array}$
\end{center}
\caption{Performance comparison of Oracle GDCS, GDCS with the proposed algorithm, the existing DCS and separate recovery when (a) ${K_{{i_j}}} = 4,\;{K_C} = 6,\;{K_{{C_{{\Pi _1}}}}} = 6,\;{K_{{C_{{\Pi _2}}}}} = 6,\;\left| {{\Pi _1}} \right| = 7,\;\left| {{\Pi _2}} \right| = 6$, (b) ${K_{{i_j}}} = 4,\;{K_C} = 5,\;{K_{{C_{{\Pi _1}}}}} = 3,\;{K_{{C_{{\Pi _2}}}}} = 3,\;\left| {{\Pi _1}} \right| = 7,\;\left| {{\Pi _2}} \right| = 6$} \label{fig4}
\end{figure}



In this section, we demonstrate the GDCS through numerical experiments.
Assuming various inter-signal correlations, we compare 
the detection performance when using Oracle-GDCS, which means GDCS with a priori-knowledge of correlation structure, 
GDCS with the proposed sequential correlation search algorithm, the existing DCS \cite{Baron:2009vd} and separate recovery.

The simulation environment is as follows.
Each signal element is generated by an i.i.d. standard Gaussian distribution, and the supports
are chosen randomly.
The signal size $N$ and the number of sensors $J$ are fixed to 50 and 9, respectively. 
As aforementioned, the identity matrix is used as a sparse basis without loss of generality.
The measurement matrix is composed of i.i.d. Gaussian entries with a variance $1/M$. 
We assume a noiseless condition in all simulations. 
The type of inter-signal correlation, the correlation structure (the number of sensors that measure partial common information) 
and the sparsity of the information
are determined as simulation parameters, and the corresponding sensors
involved in the correlation are chosen randomly.
%

We use MATLAB as a simulation tool, and YALL1 solver is used for solving the weighted ${l_1}$ minimization.
We use an iterative weighted ${l_1}$ minimization method introduced in \cite{Candes:2008hh}
to obtain adequate weight matrices within a reasonable time.
The probability of estimation error within the resolution is 
used as a performance measure where error is calculated by ${\left\| {{\bf{X}} - {\bf{\hat X}}} \right\|_2}/{\left\| {\bf{X}} \right\|_2}$, 
and the resolution is set to 0.1.

In Fig. \ref{fig3}-(a) and (b), GDCS with the proposed sequential correlation search 
outperforms the existing DCS when there exists single partial common information.
Using the GDCS with the proposed sequential correlation search, the performance is almost the same as in the oracle case.
According to the simulation, about 7 measurements are saved in (a) and about 5 measurements are saved in (b) 
comparing to GDCS with the proposed sequential correlation search and the existing DCS \cite{Baron:2009vd}. This implies that we can obtain approximately 23\% in (a) and 18\% in (b) gains
in the number of measurements when we use GDCS with the proposed sequential correlation search.
The gap between (a) and (b) is caused by different partial common information sparsity environments.

We calculate the consumed CPU time when the individual number of measurements are 25, 30, and 35 each.
The CPU time is counted only in the DCS and the GDCS with the proposed algorithm.
We average the CPU time over 100 different realizations with the simulation setting associated with Fig. \ref{fig3}-(a) and (b) each.
The units of CPU time are seconds.
In the (a) environment, the DCS CPU times are 1.42, 1.33, 1.32, respectively, while
the GDCS with the proposed algorithm CPU times are 4.07, 3.92, 3.89, respectively.
In the (b) environment, the DCS CPU times are 1.10, 1.08, 1.12, respectively, while
the GDCS with the proposed algorithm CPU times are 3.52, 3.30, 3.36, respectively.
Differently from the big-O comparison which considers the worst case,
we can observe that significant performance improvement can be achieved 
with reasonable time.


In Fig. \ref{fig4}-(a) and (b), which consider multiple inter-signal correlation including full common information and
two kinds of partial common information,
GDCS with the proposed sequential correlation search
still provides a better performance than the existing DCS \cite{Baron:2009vd}. 
However, unlike Fig. \ref{fig3}, there is a performance gap between 
Oracle-GDCS and GDCS with the proposed sequential correlation search. 
We conjecture that, 
since partial common information behaves as interference to other partial common information
during a sequential correlation search,
the algorithm fails to identify the correct correlation structure even though there is no noise.
In fact, several missed detections and incorrect detections of correlation 
are observed in the proposed algorithm in some realizations. 
About 10 measurements are saved in (a), and about 4 measurements are saved in (b) 
when we compare GDCS with with the proposed sequential correlation search and the existing DCS \cite{Baron:2009vd}. 
In total, 
26\% and 13\% gains in the number of measurements can be obtained in (a) and (b), respectively.
We also calculate the consumed CPU time when the individual number of measurements are 25, 30, 35, 40, and 45.
We average the CPU time over 100 different realizations with the simulation setting associated with Fig. \ref{fig4}-(a) and (b) each.
In the (a) environment, the DCS CPU times are 2.15, 1.98, 1.86, 1.86, 1.87, respectively, while
the GDCS with the proposed algorithm CPU times are 34.17, 34.06, 28.92, 31.97, 29.79, respectively.
In the (b) environment, the DCS CPU times are 1.42, 1.35, 1.40, 1.37, 1.36, respectively, while
the GDCS with the proposed algorithm CPU times are 16.41, 16.43, 17.39, 16.56, 15.69, respectively.
It is observed that the CPU time of the GDCS with the proposed algorithm 
increases significantly as the correlation structure becomes more complex.

\section{Conclusions}


In this paper, we extended the framework introduced in \cite{Baron:2009vd} to a realistic environment.
In the existing DCS model, partial common information must be considered as innovation information,
which cannot be used in joint recovery.
The proposed GDCS model refines the existing model so that 
it can use partial common information in the joint recovery process.
In this notion, we proposed a framework of GDCS and obtained the theoretical bound
on the number of measurements.
We also proposed a detection algorithm to identify the correlation structure so that it can exploit this information in joint signal recovery without a priori-knowledge. Numerical simulation verifies that the proposed algorithm can reduce the required number of measurements compared to the DCS algorithm. 


Future research should address a method of achieving performance close to that of
Oracle-GDCS in the presence of a large number of different partial common information. 


\appendix
\section*{Proof of Theorem \ref{thm3}}

In the appendix, we prove Theorem \ref{thm3}, the theoretical bound of the number of measurements, the general version.
Even though this proof follows the Proof of Theorem \ref{thm3} in \cite{Baron:2009vd}, 
since the considered system setup involves partial common information, the proof is nontrivial and more complex.

At first, we assume that there are three kinds of information, which are full common information, partial common information
and innovation information. Furthermore, arbitrarily, $\lambda $ kinds of partial common information exist.
Then we can write the matrix $\tilde \Phi {\bf{P}}$ as follows.

\begin{equation} \label{eq49}
\tilde \Phi {\bf{P}} = \left[ {\begin{array}{*{20}{c}}
{{\Phi _1}{P_C}}\\
{{\Phi _2}{P_C}}\\
{{\Phi _3}{P_C}}\\
 \vdots \\
{{\Phi _J}{P_C}}
\end{array}\left| {\overbrace {\begin{array}{*{20}{c}}
{{\Phi _{vec\_{C_{{\Pi _1}}}}}{P_{{C_{{\Pi _1}}}}}}& \cdots &{{\Phi _{vec\_{C_{{\Pi _\lambda }}}}}{P_{{C_{{\Pi _\lambda }}}}}}
\end{array}}^\lambda \left| {\begin{array}{*{20}{c}}
{{\Phi _1}{P_{{i_1}}}}&0&0& \cdots &0\\
0&{{\Phi _2}{P_{{i_2}}}}&0& \cdots &0\\
0&0&{{\Phi _3}{P_{{i_3}}}}& \cdots &0\\
 \vdots & \vdots & \vdots & \ddots & \vdots \\
0&0&0& \cdots &{{\Phi _J}{P_{{i_J}}}}
\end{array}} \right.} \right.} \right]
\end{equation}
where ${\Phi _{vec\_{C_{{\Pi _j}}}}}$, $j = 1,...,\lambda $ is a partially zeroed matrix corresponding to 
partial common information ${{\bf{z}}_{{C_{{\Pi _j}}}}}$
For example, if $\lambda  = 2$, ${C_{{\Pi _1}}} = \Lambda \backslash \left\{ 1 \right\}$ and 
${C_{{\Pi _2}}} = \Lambda \backslash \left\{ {1,2} \right\}$,

\begin{equation} \label{eq50}
\tilde \Phi {\bf{P}} = \left[ {\begin{array}{*{20}{c}}
{{\Phi _1}{P_C}}\\
{{\Phi _2}{P_C}}\\
{{\Phi _3}{P_C}}\\
 \vdots \\
{{\Phi _J}{P_C}}
\end{array}\left| {\begin{array}{*{20}{c}}
{{\Phi _1}{P_{{C_{{\Pi _1}}}}}}&0\\
0&0\\
{{\Phi _3}{P_{{C_{{\Pi _1}}}}}}&{{\Phi _3}{P_{{C_{{\Pi _2}}}}}}\\
 \vdots & \vdots \\
{{\Phi _J}{P_{{C_{{\Pi _1}}}}}}&{{\Phi _J}{P_{{C_{{\Pi _2}}}}}}
\end{array}\left| {\begin{array}{*{20}{c}}
{{\Phi _1}{P_{{i_1}}}}&0&0& \cdots &0\\
0&{{\Phi _2}{P_{{i_2}}}}&0& \cdots &0\\
0&0&{{\Phi _3}{P_{{i_3}}}}& \cdots &0\\
 \vdots & \vdots & \vdots & \ddots & \vdots \\
0&0&0& \cdots &{{\Phi _J}{P_{{i_J}}}}
\end{array}} \right.} \right.} \right]
\end{equation}

\begin{figure} [t]
\centering
\includegraphics [width=2.6in,height=2.4in]  {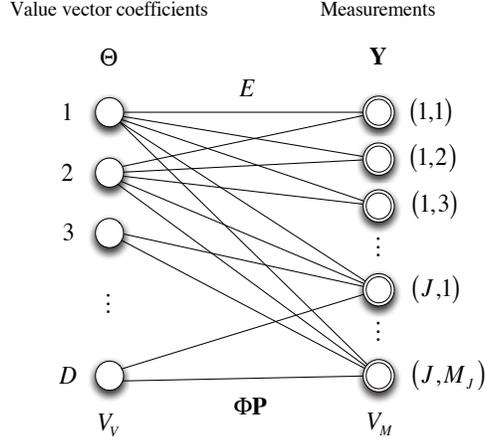}
\caption{Bipartite graph for generalized distributed compressive sensing. The graph $G = \left( {{V_V},{V_M},E} \right)$
implies the relationship between the value vector coefficients $\Theta $ and the measurements $\bf{Y}$.
} \label{fig5}
\end{figure}

Using much of \cite{Baron:2009vd}, we exploit a meaningful result from graph theory.
Before proceeding further, we define an expression for simplicity. We write ${\Pi _{include\left( j \right)}}$ 
to signify the partial common information set that includes the sensor index $j$. In other words, 
${\Pi _{include\left( j \right)}}$ represents every set among ${\Pi _1},...,{\Pi _\lambda }$ that includes $j$.
Next, a graph is considered as in Fig. \ref{fig5}.
That graph has some assumptions and properties as follows.

\begin{enumerate}[$\circ $]

\item $D$ is defined to be $D = {K_C}\left( {\bf{P}} \right) + \sum\limits_{j = 1}^\lambda  {{K_{{C_{{\Pi _j}}}}}\left( {\bf{P}} \right)}  + \sum\limits_{j \in \Lambda } {{K_{{i_j}}}\left( {\bf{P}} \right)} $, which implies the joint sparsity of the signal model.

\item For every $d \in \left\{ {1,2,...,{K_C}\left( {\bf{P}} \right)} \right\} \subseteq {V_V}$, such that column $d$ of ${P_C}$ does not also appear as a column of ${P_j}$ or of ${P_{{C_{{\Pi _{include\left( j \right)}}}}}}$, we have an edge connecting $d$ to each vertex $\left( {j,m} \right) \in {V_M}$ for $1 \le m \le {M_j}$.

\item For every $d \in \left\{ {{K_C\left( {\bf{P}} \right)+1},...,{K_C\left( {\bf{P}} \right)+{K_{{C_{{\Pi _1}}}}}\left( {\bf{P}} \right)}} \right\} \subseteq {V_V}$, such that column $d$ of ${P_{{C_{{\Pi _1}}}}}$ does not also appear as a column of of ${P_j}$
or of ${P_{{C_{{\Pi _{include\left( j \right)}} - {\Pi _1}}}}}$, we have an edge connecting $d$ to each vertex $\left( {j,m} \right) \in {V_M}$ for $1 \le m \le {M_j}$.

\item For every $d \in \left\{ {{K_C}\left( {\bf{P}} \right) + \sum\limits_{l = 1}^{k-1} {{K_{{C_{{\Pi _{l }}}}}}\left( {\bf{P}} \right)}  + 1,...,{K_C}\left( {\bf{P}} \right) + \sum\limits_{l = 1}^k {{K_{{C_{{\Pi _l}}}}}\left( {\bf{P}} \right)} } \right\} \subseteq {V_V}$, $1 < k\le\lambda$, such that column $d$ of 
${P_{{C_{{\Pi _k}}}}}$ does not also appear as a column of ${P_j}$
or of ${P_{{C_{{\Pi _{include\left( j \right)}} - {\Pi _k}}}}}$, we have an edge connecting $d$ to each vertex $\left( {j,m} \right) \in {V_M}$ for $1 \le m \le {M_j}$.

\item For every $d \in \left\{ {{K_C}\left( {\bf{P}} \right) + \sum\limits_{l = 1}^\lambda  {{K_{{C_{{\Pi _l}}}}}\left( {\bf{P}} \right)}  + 1,...,D} \right\}$, $d$ is measurable at the sensor $j$ as innovation information, and we have an edge connecting $d$ to each vertex  $\left( {j,m} \right) \in {V_M}$ for $1 \le m \le {M_j}$.

\end{enumerate}

For example, assuming that ${K_C}\left( {\bf{P}} \right) > 3$, we can observe in Fig. \ref{fig5} that edge 3 has no connection to 
$\left( {1,m} \right)$, $1 \le m \le {M_j}$. This is because there is an overlap between full common information and other information at the sensor 1. 

Now we are ready to exploit Hall's marriage theorem for bipartite graphs \cite{West.IntroductiontoGraph}.
Hall's marriage theorem states that, assuming an arbitrarily bipartite graph $\left( {{V_1},{V_2},E} \right)$, 
if the cardinality of an arbitrary subset of ${V_1}$ is always greater than the cardinality of its neighbors in ${V_2}$,
every element of ${V_1}$ has a unique neighbor in ${V_2}$ with no overlap.

In the graph $G = \left( {{V_V},{V_M},E} \right)$, let us consider an arbitrary subset $\pi $, and 
let $E\left( \pi  \right)$ denote the set of neighbors in ${V_M}$.
To satisfy Hall's marriage theorem,
$\left| {E\left( \pi  \right)} \right| \ge \left| \pi  \right|$ for any $\pi $.
If we let ${S_\pi } = \left\{ {j \in \Lambda \left| {\left( {j,m} \right) \in E\left( \pi  \right){\rm{ \;for\;some\;}}m} \right.} \right\}$,
the following statement is required for assigning a unique neighbor to every edge in ${V_V}$.

\begin{equation} \label{eq51}
\sum\limits_{j \in {S_\pi }} {{M_j}}  \ge \left| \pi  \right|
\end{equation}
Therefore, it would be sufficient to prove the following inequality.

\begin{equation} \label{eq52}
{O_C}\left( {{S_\pi },{\bf{P}}} \right) + \sum\limits_{{{\bf{z}}_{{C_\Pi }}} \in {\Omega _2}\left( {{S_\pi }} \right)} {{O_{{C_\Pi }}}\left( {{S_\pi },{\bf{P}}} \right)}  + \sum\limits_{{{\bf{z}}_{{C_\Pi }}} \in {\Omega _1}\left( {{S_\pi }} \right)} {{K_{{C_\Pi }}}\left( {\bf{P}} \right)}  + \sum\limits_{j \in {S_\pi }} {{K_{{i_j}}}\left( {\bf{P}} \right)}  \ge \left| \pi  \right|
\end{equation}
Then, since the condition of Theorem \ref{thm3} includes all subsets $\Gamma  \in \Lambda $, Hall's marriage theorem 
is satisfied for the graph $G$. 
To prove (\ref{eq50}), we divide $\pi $ into three disjoint parts, i.e., 
$\pi  = {\pi _I} \cup {\pi _{{PC}}} \cup {\pi _C}$.

First, as shown in \cite{Baron:2009vd}, 
$\left| {{\pi _I}} \right| \le \sum\limits_{j \in {S_\pi }} {{K_{{i_j}}}\left( {\bf{P}} \right)} $ since
all innovation information measured by ${S_\pi }$ are counted.
Next, we claim that $\left| {{\pi _C}} \right| \le {O_C}\left( {{S_\pi },{\bf{P}}} \right)$.
This is because ${\pi _C} \subseteq \pi $, and it implies that 
${O_C}\left( {{S_{{\pi _C}}},{\bf{P}}} \right) \le {O_C}\left( {{S_\pi },{\bf{P}}} \right)$.
Consequently, $\left| {{\pi _C}} \right| \le {O_C}\left( {{S_\pi },{\bf{P}}} \right)$.
When we consider ${\pi _{PC}}$, 
we write 
${\pi _{PC}} = {\pi _{PC,\;{\Omega _1}\left( {{S_\pi }} \right)}} \cup {\pi _{PC,\;{\Omega _2}\left( {{S_\pi }} \right)}}$,
where ${\pi _{PC,\;{\Omega _j}\left( {{S_\pi }} \right)}}$ contains partial common information included 
in ${\pi _{PC}}$ and ${\Omega _j}\left( {{S_\pi }} \right)$ simultaneously.
Here, $j = 1,\;2$.
In terms of ${\pi _{PC,\;{\Omega _1}\left( {{S_\pi }} \right)}}$, by the same reason of the case of innovation information,
$\left| {{\pi _{PC,\;{\Omega _1}\left( {{S_\pi }} \right)}}} \right| \le \sum\limits_{{{\bf{z}}_{{C_\Pi }}} \in {\Omega _1}\left( {{S_\pi }} \right)} {{K_{{C_\Pi }}}\left( {\bf{P}} \right)} $. When ${\pi _{PC,\;{\Omega _2}\left( {{S_\pi }} \right)}}$ is considered, 
by the same reason of the case of full common information, 
$\left| {{\pi _{PC,\;{\Omega _2}\left( {{S_\pi }} \right)}}} \right| \le \sum\limits_{{{\bf{z}}_{{C_\Pi }}} \in {\Omega _2}\left( {{S_\pi }} \right)} {{O_{{C_\Pi }}}\left( {{S_\pi },{\bf{P}}} \right)} $.
Considering all of this, we claim that, if Theorem \ref{thm3} is satisfied, 
every element in ${V_V}$ has an unique matching in ${V_M}$.

Before proceeding to the next step, we introduce an important lemma.

\newtheorem{lemma1}{Lemma}

\begin{lemma1}[\cite{Baron:2009vd}, Full rank property of a Gaussian or zero vector matrix] \label{lem1}
Let ${\Upsilon ^{\left( {d - 1} \right)}}$ be a $\left( {d - 1} \right) \times \left( {d - 1} \right)$ matrix
having full rank. Construct $d \times d$ matrix ${\Upsilon ^{\left( d \right)}}$ as follows:
\begin{equation} \label{eq53}
{\Upsilon ^{\left( d \right)}}: = \left[ {\begin{array}{*{20}{c}}
{{\Upsilon ^{\left( {d - 1} \right)}}}&{{v_1}}\\
{v_2^t}&\omega 
\end{array}} \right]
\end{equation}
where ${v_1},{v_2} \in {\mathbb{R}^{d - 1}}$ are vectors with each entry being either zero or a Gaussian random variable, 
$\omega $ is a Gaussian random variable, and all random variables are i.i.d. and independent of ${\Upsilon ^{\left( d-1 \right)}}$. Then with probability one, ${\Upsilon ^{\left( d \right)}}$ has full rank.
\end{lemma1}

The next step is equivalent with the method used in \cite{Baron:2009vd}. Here, we briefly describe the process.
At first, by subtracting the columns of innovation information from the columns of full common information and partial common information, all the same valued columns are removed, and a partially zeroed matrix is formed.
Second, based on the obtained unique neighbor, the rows are chosen. For example, 
if the sensor $j$ contains $K$ neighbors, 
$K$ rows are chosen within the part of the sensor $j$, i.e., the rows of the matrix
corresponding to the sensor $j$.
Finally, in the constructed matrix that consists of chosen rows in the second step, 
the columns are rearranged.
In the order of the sensor index, 
the information component of which its neighbor is in the corresponding sensor is placed.
For example, if the sensor $j$ contains 
the neighbor of the $k$th component of full common information, the $k$th column 
of full common information part is stacked in the space of the sensor $j$.
All these operations are column subtraction and column or row choice, which means
there is no rank increase.
Then, by the construction, all the diagonal elements of the constructed matrix
are i.i.d. Gaussian random variables, and the other part is 
a zero or i.i.d. Gaussian random variable. Therefore, by Lemma \ref{lem1}, 
$\tilde \Phi {\bf{P}}$ has full rank. 






\ifCLASSOPTIONcaptionsoff
  \newpage
\fi



\bibliographystyle{IEEEtran}
\bibliography{gdcs_bib}

\end{document}